\documentclass[12pt,a4paper,english]{article}
\usepackage[T1]{fontenc}
\usepackage[latin9]{inputenc}
\usepackage{array}
\usepackage{multirow}
\usepackage{amsmath}
\usepackage{amssymb}
\usepackage{graphicx}
\usepackage{esint}
\usepackage{lmodern}

\makeatletter

\providecommand{\tabularnewline}{\\}

\@ifundefined{extrarowheight}
{\usepackage{array}}{}
\renewcommand{\vec}[1]{{|\,#1\,\rangle}}

\newcommand{\cH}{{\cal H}}

\makeatother

\usepackage{babel}

\usepackage[top=2.5cm, bottom=2.5cm, left=2.5cm, right=2.5cm]{geometry}
\usepackage[width=\textwidth]{caption}

\begin{document}

\title{Defect scaling Lee-Yang model from the perturbed DCFT point of view}

\maketitle

\begin{center}
Zoltán Bajnok\textsuperscript{a}, László Holló\textsuperscript{a} and Gerard Watts\textsuperscript{b}
\end{center}

\begin{center}
\textsuperscript{a}MTA Lendület Holographic QFT Group,

Wigner Research Centre for Physics

H-1525 Budapest 114, P.O.B. 49, Hungary 

\textsuperscript{b}Department of Mathematics, King\textquoteright{}s College
London,

Strand, London WC2R 2LS \textendash{} UK
\end{center}

\begin{abstract}
We analyze the defect scaling Lee-Yang model from the perturbed defect
conformal field theory (DCFT) point of view. First the defect Lee-Yang
model is solved by calculating its structure constants from the sewing
relations. Integrable defect perturbations are identified in conformal
defect perturbation theory. Then pure defect flows connecting integrable
conformal defects are described. We develop a defect truncated conformal
space approach (DTCSA) to analyze the one parameter family of integrable
massive perturbations in finite volume numerically. Fusing the integrable
defect to an integrable boundary the relation between the IR and UV 
parameters can be derived from the boundary relations. We checked these 
results by comparing the spectrum for large volumes to the scattering 
theory.
\end{abstract}

\section{Introduction}
\label{sec:intro}

Recently there has been growing interest and relevant progress in
analyzing integrable defect theories. According to the no-go theorem
\cite{Delfino:1994nr} relativistic interacting integrable defect
theories are purely transmitting. Thus the analysis of such theories
concerned the construction of classical field theories admitting integrable
defects \cite{Bowcock:2003dr,Gomes:2007cc,Caudrelier:2008zz,Corrigan:2009vm,
Avan:2012rb,Doikou:2012fc}
and their exact solutions in terms of the exact transmission factors
of the particles \cite{Konik:1997gx,Bowcock:2005vs,Bajnok:2007jg,Corrigan:2010ph,
Corrigan:2010sr}. 

In the classical constructions an important observation was that fields
need not be continuous at the defect \cite{Bowcock:2003dr}. Cleverly
chosen defect potential terms ensure the conservation of higher spin
charges, which seems to be equivalent to the conservation of momentum
\cite{Corrigan:2007gt}. Interestingly the defect condition, obtained
from the variation of the action, basically implements the B\"acklund
transformation of the theory \cite{Habibullin:2007ww}. The
integrable classical defect theories constructed this way have to be
quantized, which can be implemented in various schemes. For standard
schemes, their relations and implementations see \cite{Bajnok:2012}.

There is a scheme which is based on the action: one can separate
free and perturbation parts, quantize the free part first and take
into account the perturbation iteratively. This scheme is useful to
prove exact integrability and to show that transmission factors satisfy
unitarity and crossing symmetry. On top of this it might provide a
way to calculate the transmission factors perturbatively \cite{Bajnok:2007jg}. 

In the bootstrap scheme, integrability is assumed and the transmission
factors are determined from their consistency relations, such as unitarity,
crossing symmetry, Yang-Baxter equation and maximal analyticity \cite{Bajnok:2004jd}.
(All poles of the transmission factors have to be explained by some
Coleman-Thun type diagrams.). These requirements lead to a solution
for the transmission factors, which solves the theory in infinite
volume (IR) exactly. However, this solution is not unique: it contains
CDD-type ambiguities which have to be fixed. Additionally, even the
minimal solution may contain parameters which have to be related to
that of the action. A typical way of doing this is to solve the theory
in finite volume. By varying the volume we can smoothly interpolate
between the large volume (transmission) and small volume (action)
description and connect the parameters on the two sides \cite{Bajnok:2007jg}. 

In the present paper we analyze the scaling Lee-Yang model on the
circle with integrable defect conditions. This is the simplest nontrivial integrable
scattering theory having only one type of particle. The bootstrap
solution of the model was carried out in \cite{Bajnok:2007jg}, where
a one parameter family of transmission factors were determined. Maximal
analyticity was checked by closing the defect bootstrap: all poles
of the transmission factors have been explained either by defect bound-states
or by Coleman-Thun type diagrams. The ground-state defect
thermodynamic Bethe ansatz (DTBA) equation
has been derived as well, which leads to the exact determination of
the bulk energy constant and defect energy. Our aim is to connect
this (IR) description to the one based on the action of the model.
In doing so we have to determine the UV conformal field theory appearing
in the small volume limit. Then we have to identify its integrable
perturbations, and finally relate the parameters appearing in the
UV and IR descriptions. We achieve these aims as follows: 

In section \ref{sec:dlym} we recall the topological defects of the
Lee-Yang model  
and solve them completely by calculating all relevant structure constants.
In section \ref{sec:mlessp} we use the perturbed defect conformal
field theory (DCFT) 
framework to identify the integrable 
perturbations localized on the defect. These are either holomorphic
or anti-holomorphic massless perturbations and induce flows between
different conformal defect conditions. In Section \ref{sec:massp}
 we introduce a
mass scale by perturbing in the bulk and analyze under what circumstances
the combined bulk and defect perturbations preserve momentum. We find
a one parameter family of integrable perturbations just as we have
in the IR. In section \ref{sec:dtcsa} we develop a defect truncated
conformal space 
approach (DTCSA) method to analyze
the spectrum in finite volume numerically. We recall the IR description
of the model in Section \ref{sec:6} and compare the numerical DTCSA data to
the defect Bethe-Yang equations in \ref{sec:numer}. We establish the
UV-IR relation and 
comment on how the defect results are related to the boundary ones.
Finally we conclude in Section 8. The technical details of calculating
the structure constants and performing perturbed DCFT calculations
are relegated to Appendices.

\section{Defect Lee-Yang model}
\label{sec:dlym}

The Lee-Yang model is the simplest nontrivial conformal field theory.
It has central charge $c=-\frac{22}{5}$ and the Virasoro algebra
contains just two irreducible modules with highest weights $0$ and
$-\frac{1}{5}$, respectively. The identity module, $V_{0}$, is built
over the $sl_{2}$ invariant vacuum $\vert0\rangle$ as \cite{Feigin:1991wv,Fortin:2004ct}
\begin{equation}
L_{-n_{1}}\dots L_{-n_{m}}\vert0\rangle\quad;\qquad n_{m}>1\qquad;\qquad n_{i}>n_{i+1}+1
\end{equation}
Interestingly, this basis is linearly independent (there are no singular
vectors) and it is related to the fermionic type reduced character
\cite{Nahm:1992sx}:
\begin{equation}
\chi_{0}(q)=\sum_{n=1}^{\infty}\mbox{dim}(V_{0}^{(n)})q^{n}=
\sum_{n=1}^{\infty}\frac{q^{n^{2}+n}}{(1-q)\dots(1-q^{n})}=
\prod_{n=1}^{\infty}\frac{1}{(1-q^{5n-3})(1-q^{5n-2})}
\end{equation}
where $V_{0}^{(n)}$ denotes the level $n$ subspace of $V_{0}$. 

The other module appearing, $V_{1},$ is built over the highest weight
state $\vert h\rangle$ where here and from now on $h=-\frac{1}{5}$.
The module is generated by the modes 
\begin{equation}
L_{-n_{1}}\dots L_{-n_{m}}\vert h\rangle\quad;\qquad n_{m}>0\qquad;\qquad n_{i}>n_{i+1}+1
\end{equation}
 and has the reduced character: 
\begin{equation}
\chi_{1}(q)=\sum_{n=1}^{\infty}\mbox{dim}(V_{n}^{1})q^{n}=
\sum_{n=1}^{\infty}\frac{q^{n^{2}}}{(1-q)\dots(1-q^{n})}=
\prod_{n=1}^{\infty}\frac{1}{(1-q^{5n-4})(1-q^{5n-1})}
\end{equation}
The fusion relations of the model can be encoded into 
$N_{00}^{0}=N_{01}^{1}=N_{10}^{1}=N_{11}^{0}=N_{11}^{1}=1$
and $N_{00}^{1}=N_{01}^{0}=N_{10}^{0}=0$. 

The Lee-Yang model with periodic boundary condition carries a representation
of \hbox{$\hbox{\em Vir}\otimes\overline{\hbox{\em Vir}}$} and its Hilbert space can be decomposed
as 
\begin{equation}
\mathcal{H}=V_{0}\otimes\bar{V}_{0}+V_{1}\otimes\bar{V}_{1}\label{eq:bulkH}
\end{equation}
 For each vector of the Hilbert space there is an associated local
field; in particular $\vert0,0\rangle\to\mathbb{I}$ and $\vert
h,h\rangle\to\Phi(z,\bar{z})$. 
These fields form an operator algebra, with structure constants
\begin{equation}
  \Phi(z,\bar{z})\Phi(0,0)
= \frac{C_{\Phi\Phi}^{\mathbb{I}}\mathbb{I}}{\vert z\vert^{4h}}
+ C_{\Phi\Phi}^{\Phi}\frac{\Phi(0,0)}{\vert z\vert^{2h}}
+ \dots
\end{equation}
One consistent choice of these constants is
\begin{equation}
  C_{\Phi\Phi}^{\mathbb{I}}
= -1
\quad;\quad 
  C_{\Phi\Phi}^{\Phi}
= \sqrt{\frac{2}{1+\sqrt{5}}}
  \frac{\Gamma(\frac{1}{5})\Gamma(\frac{6}{5})}
       {\Gamma(\frac{3}{5})\Gamma(\frac{4}{5})}
= \,1.9113127..\label{eq:Bulkstructureconstants}
\end{equation}
Although the normalization $C_{\Phi\Phi}^{\mathbb{I}}=-1$ seems a
bit unnatural, it is a consequence of the fact that the Lee-Yang model
is non-unitary and our insistence on having a real field
\begin{equation}
\Phi^{\dagger}=\Phi
\end{equation}
Had we chosen $C_{\Phi\Phi}^{\mathbb{I}}=1$ it would lead to $\Phi^{\dagger}=-\Phi$
and imaginary $C_{\Phi\Phi}^{\Phi}.$ The three point coupling $C_{\Phi\Phi}^{\Phi}$
can be determined by requiring the single-valuedness of the four point
functions. See Appendix A for the details. The structure constants
of descendant operators follow from consistency requirements of the
Virasoro algebra. 

From now on we analyze the Euclidean theory on the finite cylinder
$z=x+iy$ with periodic space coordinates $x\equiv x+L$, and with
Euclidean time $y=it$. Taking the limit $L\to\infty$ we obtain a
theory on the plane, which is useful to analyze local properties such
as conservation laws. 

The energy momentum tensor on the plane has a holomorphic $T(z)$
and an anti-holo\-morphic component, $\bar{T}(\bar{z})$, whose modes
generate two commuting Virasoro algebras, the symmetry algebra
of the theory. The conservation laws 
\begin{equation}
\partial_{\bar{z}}T(z)=0\quad;\qquad\partial_{z}\bar{T}(\bar{z})=0\label{eq:Tconslaw}
\end{equation}
 lead to the conservation of energy and momentum\footnote{Since we choose the complex coordinates as $z=x+i y$, this momentum generates the shifts to the negative $x$ directions.} 
\begin{equation}
H=\int dx \, (T(z)+\bar{T}(\bar{z}))\quad;\qquad P=\int dx\, (T(z)-\bar{T}(\bar{z}))
\end{equation}
which is a consequence of the fact that fields vanish at spacelike
infinities: 
\begin{equation}
\partial_{y}H=i\int dx \, \partial_{x}(T(z)-\bar{T}(\bar{z}))=0\quad;
\qquad\partial_{y}P=i\int dx\, \partial_{x}(T(z)+\bar{T}(\bar{z}))=0
\end{equation}

We introduce a conformally invariant defect condition at the line $x=0$
by demanding the conservation of energy: the energy flow leaving the
left part has to appear on the right part. Denoting the fields living
on the left part, ($x<0$), as $T_{-},\bar{T}_{-}$, while the ones
on the right, ($x>0$), by $T_{+},\bar{T}_{+}$ the energy can be
written as the sum of the energies of the two parts: 
\begin{equation}
  H
= H_{-}+H_{+}
= \int_{-\infty}^{0}dx \,
  (T_{-}(z)+\bar{T}_{-}(\bar{z}))
+ \int_{0}^{\infty}dx \,
  (T_{+}(z)+\bar{T}_{+}(\bar{z}))
\label{eq:Hp+Hm}
\end{equation}
As the conservation laws (\ref{eq:Tconslaw}) in the bulk, $(x\neq0$),
are not affected by the presence of the defect, the conservation of
the total energy, $\partial_{y}H=0$, gives the constraint 
\begin{equation}
\partial_{y}H=i\lim_{x\to-0}(T_{-}(z)-\bar{T}_{-}(\bar{z}))-i\lim_{x\to+0}
(T_{+}(z)-\bar{T}_{+}(\bar{z}))=0
\;.
\end{equation}
Defects which preserve the energy are called {\em conformal} defects
and the Lee-Yang model is unique for having the possible conformal
defects completely classified \cite{Quella:2006de}.

Similarly, demanding the conservation of total momentum 
\begin{equation}
  P
= P_{-}+P_{+}
= \int_{-\infty}^{0}dx\, (T_{-}(z)-\bar{T}_{-}(\bar{z}))
+ \int_{0}^{\infty}dx\, (T_{+}(z)-\bar{T}_{+}(\bar{z}))
\label{eq:Pp+Pm}
\end{equation}
we obtain the condition 
\begin{equation}
\partial_{y}P=i\lim_{x\to-0}(T_{-}(z)+\bar{T}_{-}(\bar{z}))-i\lim_{x\to+0}
(T_{+}(z)+\bar{T}_{+}(\bar{z}))=0
\end{equation}
which, when combined with the energy conservation, leads to the separate
conservation of the holomorphic and anti holomorphic parts:
\begin{equation}
\lim_{x\to-0}T_{-}(z)=\lim_{x\to+0}T_{+}(z)\quad;\qquad\lim_{x\to-0}\bar{T}_{-}
(\bar{z})=\lim_{z\to+0}\bar{T}_{+}(\bar{z})
\label{eq:cont}
\end{equation}
 Such defects, preserving both energy and momentum, are unseen by
the energy momentum tensor. They are called \emph{topological} or
\emph{purely transmitting} defects. 

Other fields, say $\Psi$,  can see the defect since in general they
are not necessarily continuous there: 
\begin{equation}
\lim_{x\to-0}\Psi_{-}(z,\bar{z})\neq\lim_{x\to+0}\Psi_{+}(z,\bar{z})
\end{equation}
where we denote the bulk field living on the left/right
part of the defect by $\Psi_{\mp}(z,\bar{z})$. Nevertheless, both
can be expanded in terms of defect fields via the bulk-defect OPE\footnote{Note that this form of the OPE is true when the defect is oriented vertically}:
\begin{equation}
  \Psi_{\mp}^{i}(z,\bar{z})
= \sum_{j}
  C_{\Psi_{\mp},j}^{i}
  \vert x\vert^{h_{i}-h_{j}}
  \vert x\vert^{\bar{h}_{i}-\bar{h}_{j}}
  \psi^{j}(y)
\label{eq:bdope}
\end{equation}
where $\psi^{j}(y)$ transforms covariantly under the two copies
of the Virasoro algebras corresponding to $T(y)$ and
$\bar{T}(y)$. \cite{Petkova:2000ip,Quella:2006de}.  

\begin{figure}
\begin{centering}
\includegraphics[bb = 0 0 525 235, scale=0.7, type=eps]{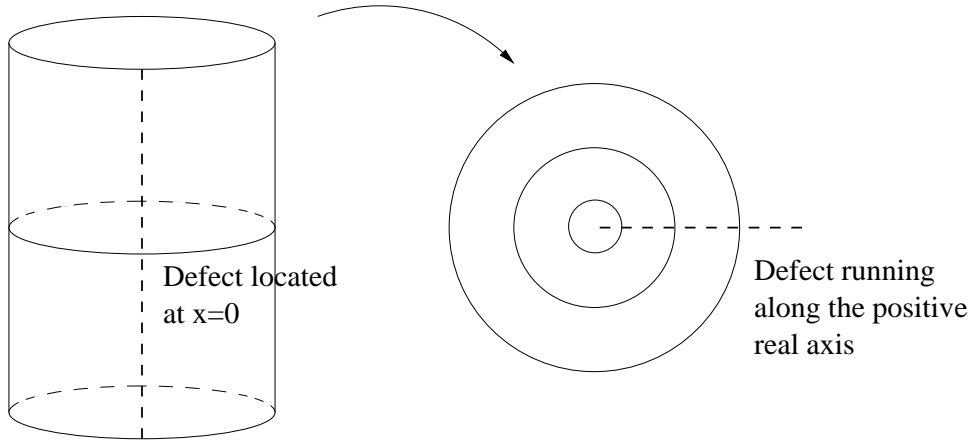}
\par\end{centering}

\caption{A defect placed at $x{=}0$ on the infinite cylinder is mapped to
  the positive real axis; 
  states in the Hilbert space corresponds
  to fields located at the origin of the plane, that is fields living
  at.the end of the defect.}

\label{fig:pic}
\end{figure}

So far we considered defects as located at the point $x{=}0$ on the real
line. It is more usual in conformal field theory to take space to be
compact, so that the defect can be considered as running along an
infinite cylinder $0\leq x < 2 \pi$ with the Hamiltonian generating
translations along the cylinder. The defect is again located at $x{=}0$.
The cylinder can be conformally mapped to the plane by 
$z \mapsto \exp(i z)$, with constant time slices being circles of
constant radius and the defect now 
running along the positive real axis, as in figure \ref{fig:pic}.
Since a topological defect is invisible to the
holomorphic and anti-holomorphic parts of the energy momentum tensor,
the Hilbert space of this system carries the action of two commuting
copies of the Virasoro algebra, and can be decomposed into sums of
pairs of representations of the algebra. 
With the end of the defect now located at the origin of the complex
plane, this Hilbert space corresponds to the fields that can live at
the end of a defect, which one can think of as defect-creating or
defect-annihilating fields.

\newcommand{\blank}[1]{}
\blank{Topological defects can be considered, alternatively, as operators acting
on the bulk Hilbert space (\ref{eq:bulkH}) commuting with the action
of the left and right Virasoro generators \cite{Petkova:2000ip,Quella:2006de}.
They have to act diagonally on each factor in (\ref{eq:bulkH}) and
satisfy a Cardy type condition. The Cardy type condition formulates
the covariance of the torus partition function under modular
transformations. When the defect acts as an operator on the periodic
Hilbert space the partition function is the weighted sum of the
diagonally combined left and right Virasoro characters. Alternatively,
when it is located in space it modifies the Hilbert space. As the
defect is conformal this modified space carries a representation of
the tensor product of the left and right Virasoro algebras. Evaluating
the partition function as a sum over these (generally not left/right
symmetric)  representation spaces and equating with the modularly
transformed calculations provides severe 
restriction both on the defect Hilbert space and the defect operator. In
the Lee-Yang theory these lead to two choices for purely
transmitting defects, which can be labelled
in the same way as we label bulk fields: $(0,0)$ and $(1,1)$.}

The identity defect, $(0,0)$, is invisible
to all fields, in other words it is the formal solution of the defect
equations (\ref{eq:cont}) which corresponds to the absence of any defect
whatsoever. This means that the Hilbert space containing the defect 
is the same as the Hilbert space in the absence of a defect - 
\begin{equation}
\mathcal{H}^{(0,0)}=V_{0}\otimes\bar{V}_{0}+V_{1}\otimes\bar{V}_{1}
\;,
\end{equation}
and the corresponding highest weight states are
$\vec{0,0} = \vec 0$ and $\vec{h,h}=\vec{\Phi}$.
Furthermore, the operators living on the defect are just the bulk
fields with the same operator algebra.

In the case of the $(1,1)$ defect, the Hilbert space decomposes as
\cite{Petkova:2000ip} 
\begin{equation}
\mathcal{H}^{d}=V_{1}\otimes \bar{V}_{0}+V_{0}\otimes \bar{V}_{1}
+V_{1}\otimes \bar{V}_{1}
\end{equation}
The corresponding highest weight states will be denoted as 
\begin{equation}
\vert d\rangle=\vert h,0\rangle\quad;\qquad\vert\bar{d}\rangle=\vert0,h
\rangle\quad;\qquad\vert D\rangle=\vert h,h\rangle
\end{equation}
and they form (up to signs) an ortho-normal basis. 
These highest weight states correspond to fields which can be located
at the end of the defect, or ``create'' the defect, as in figure
\ref{fig:pic}. 
Similarly to the
bulk normalization 
eq. (\ref{eq:Bulkstructureconstants}) we choose the normalization
$\langle D\vert D\rangle=-1$. 

\begin{figure}
\begin{centering}
\includegraphics[bb = 0 0 525 235, scale=0.7, type=eps]{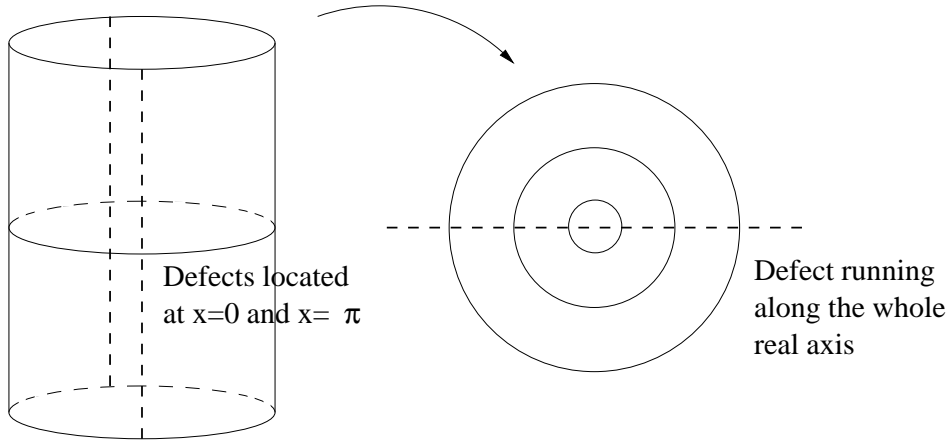}
\par\end{centering}

\caption{Two defects placed at $x{=}0$ and $x=\pi$ on the infinite
  cylinder are mapped to a single defect running along the whole real
  axis on the complex plane; states in the Hilbert space corresponds
  to fields located at the origin of the plane, that is fields living
  on the defect.}

\label{fig:pic2}
\end{figure}

The operators living (not at the end but) on the defect correspond to
the Hilbert space of the model on a circle where the defect runs
along the whole real line, so that fields at the origin are located on
the defect, as in figure \ref{fig:pic2}. When mapped back to the
cylinder, this corresponds to a 
cylinder with {\em two} defects \cite{Petkova:2000ip}: 
\begin{equation}
\mathcal{H}^{(1,1)}=V_{0}\otimes\bar{V}_{0}+V_{1}\otimes\bar{V}_{0}+
V_{0}\otimes\bar{V}_{1}+2\, V_{1}\otimes\bar{V}_{1}
\end{equation}
The h.w. vectors of these representation spaces we correspond to the following primary defect fields $\mathbb{I}$, $\varphi$, $\bar{\varphi}$, $\Phi_{-}$, $\Phi_{+}$, respectively.
The non-chiral fields $\Phi_{\pm}$ can be taken to be the left/right
limits of the bulk field $\Phi(z,\bar{z})$ on the defect. We calculate
the structure constants of this defect conformal field theory in Appendix
A by solving the sewing relations. 

\section{Massless perturbations}
\label{sec:mlessp}

In this section we search for massless perturbations of the defect Lee-Yang
model that preserve both energy and momentum. First we introduce a
chiral holomorphic perturbation, then an anti-holomorphic one, finally
we analyze a combination of chiral and anti-chiral perturbations.

\subsubsection*{Holomorphic perturbation}

We consider a chiral holomorphic perturbation of the form
\begin{equation}
S=S_{DCFT}-\mu\int_{-\infty}^{\infty}dy\,\varphi(y) 
\label{eq:pact}
\end{equation}
 As the action is dimensionless and the primary field has dimension
$[\varphi]=h$ the dimension of the coupling constant is $[\mu]=1-h$.
The perturbation on the defect does not affect the conservation laws
in the bulk but it may change the bulk-defect OPE 
(\ref{eq:bdope}) or, consequently, the defect condition
(\ref{eq:cont}). 

The change of the defect condition has a series expansion in $\mu$
which we can calculate in perturbation theory:
\begin{equation}
  \Delta T(y):=T_-(y)-T_+(y)
=\mu\mathcal{O}_{1}(y)
+\mu^{2}\mathcal{O}_{2}(y)
+\dots+\mu^{n}\mathcal{O}_{n}(y)+\dots 
\end{equation}
where $\mathcal{O}_{n}$ are operators localized on the defect. Each operator
equation is understood within correlators in the perturbed theory, for
renormalized operators. Comparing the dimensions of the two sides we
observe that the dimension of the 
operator appearing, $\mathcal{O}_{n}$, has to be 
\begin{equation}
[\mathcal{O}_{n}]=nh-n+2
\end{equation}
As the most negative left chiral dimension is $-\frac{1}{5}$ the
only non-vanishing contribution can appear for $n=1$ with
$\mathcal{O}_{1}\propto\partial\varphi$.

As a consequence, the corresponding change in the bulk-defect OPE must have
the form
\begin{equation}
 T(x+iy) = \begin{cases}
T^R(y) + b^+\,\mu\,\partial_y{\varphi}(y) + O(x) \;,& x>0 \\
T^R(y) + b^-\,\mu\,\partial_y{\varphi}(y) + O(x) \;,& x<0 
\end{cases}\;,
\end{equation}
where $T^R(y)$ is a suitably renormalised field and $b^\pm$ are
constants. 

However, the calculation of $b^\pm$  depends in detail on the regulation of
divergences in the perturbation expansion and the precise definitions
of the bulk and defect fields in the perturbed theory. 
We shall regulate the perturbation expansion using a hard cut-off
$\epsilon$, so that in (\ref{eq:pact}) the integration is only over
values $y$ such that $|y-z|>\epsilon$ where $z$ is the insertion point
of any other local field, either bulk or defect. 

When the field $T(x+iy)$ approaches to within a
distance $\epsilon$ of the defect, 
because the perturbation is cut off at distances
less than $\epsilon$
the effect of the perturbation is
reduced. As $x\to 0$, with $\epsilon$ fixed, the defect appears
unperturbed and the structure constants $b^\pm$ go to zero. 
It is possible to keep careful track throughout our calculations of
whether fields approach closer than $\epsilon$ to a defect, but to
simplify the discussion we shall always assume that the limit
$\epsilon\to 0$ is taken {\em before} any other limits. With this
assumption, we find that $b^\pm = \mp i \pi(1-h)$ (see appendix
\ref{app:details} for details) and so  
\begin{eqnarray}
\Delta T(y)
&:=&
\lim_{x\to0^{+}}(T_{-}(-x+iy)-T_{+}(x+iy))
\nonumber\\
&=& 
\lim_{x\to0^{+}} \lim_{\epsilon\to 0} (T_{-}(-x+iy)-T_{+}(x+iy))  
\nonumber\\
&=&
 2\pi i(1-h)\mu \,\partial_y \varphi(y)
\;.
\end{eqnarray}
The anti-holomorphic part is not changed, $\Delta\bar{T}=0$. 

This first order perturbative result is exact to any order in $\mu$\footnote{Here and from now on we introduce $T_{0}$ to distinguish the conformal energy-momentum tensor from the perturbed one.}:
\begin{equation}
\Delta T_{0}(y)e^{\mu\int_{-\infty}^{\infty}dy'\,\varphi(y')}=2\pi\mu(1-h)
i(\partial_{y}\varphi(y))e^{\mu\int_{-\infty}^{\infty}dy'\,\varphi(y')}
\end{equation}
where here and from now on operator products are always time ($y$) ordered, 
which we do not write out explicitly. 

As the jump of the energy momentum tensor is a total derivative we
can define the conserved energy as 
\begin{equation}
H=H_{-}+H_{+}+2\pi\mu(1-h)\varphi
\end{equation}
The existence of a conserved energy is not very surprising as our
system is invariant under time translations. What is more surprising
is that the momentum 
\begin{equation}
P=P_{-}+P_{+}+2\pi\mu(1-h)\varphi
\end{equation}
is also conserved, although we do not have translational invariance.
This also means that the defect remains topological after the perturbation. 

As there are only two topological defect conditions we expect
a defect flow from the $\mathcal{H}^{d}$ defect to the $\mathcal{H}^{(0,0)}$
identity defect as the coupling constant ${\mu}$ increases. 
If we
plot the eigenvalues of the dimensionless operator $\frac{HL}{2\pi}$,
as a function of the dimensionless parameter $\mu L^{6/5}$
we can identify the states in both Hilbert spaces as well as the flows.

Lattice calculations  
\cite{BP}
give a lot of information on these flows; in particular they describe
the whole space of flows, in the following sense. 

The UV endpoint of a flow is an energy and momentum eigenstate in the
$\cH^{d}$ space. This means it is an eigenstate of both $L_0$ and
$\bar L_0$ and so is a descendant at $L_0$--level $M$ and $\bar
L_0$--level $\bar N$ of some highest weight state in $\cH^{d}$. 
Hence the UV endpoints of the flows form a distinguished basis of
states and (from the results in section \ref{sec:dlym})
we can label them by two sets of
integers,  $\{m_i;\bar n_j\}$ satisfying $\sum_i m_i = M$ and $\sum_j
\bar n_j = \bar N$ and certain other restrictions, depending on the
sector in the Hilbert space.

Likewise, the IR endpoints of the flows determine a distinguished
basis of states in $\cH^{(0,0)}$ labelled by another two sets of
integers $\{m'_i;\bar n'_j\}$ satisfying another set of
restrictions. Since the flow is entirely holomorphic, the
anti-holomorphic representation cannot change and 
$\bar n'_j =\bar n_j$ but the lattice calculations in \cite{BP}
indicate that the holomorphic representations and the integers
$\{m'_i\}$ and $\{m_i\}$ are related as in table \ref{table:1}.

\begin{table}
{\renewcommand{\arraystretch}{1.4}\[
\begin{array}{c|rcl|c}
  \hbox{UV} 
& 
& 
& 
& \hbox{IR}
\\\hline
  (d) 
& \{m_1,\cdots,m_l ; {\bar n}_1,\cdots,{\bar n}_k\}
& \longrightarrow
& \{m_1{+}1,\cdots,m_l{+}1 ; {\bar n}_1,\cdots,{\bar n}_k\}
& (0,0)
\\
& 
\multicolumn{3}{c|}{
m_l\geq 1
\;,\;\;
{\bar n}_k\geq 2
}
\\\hline
  (\bar d) 
& \{m_1,\cdots,m_l ; {\bar n}_1,\cdots,{\bar n}_k\}
& \longrightarrow
& \{m_1{+}1,\cdots,m_l{+}1,1 ; {\bar n}_1,\cdots,{\bar n}_k\}
& (h,h)
\\ 
& 
\multicolumn{3}{c|}{
m_l\geq 2
\;,\;\;
{\bar n}_k\geq 1
}
\\ \hline
  (D) 
& \{m_1,\cdots,m_l;{\bar n}_1,\cdots,{\bar n}_k\}
& \longrightarrow
& \{m_1{+}1,\cdots,m_l{+}1;{\bar n}_1,\cdots,{\bar n}_k\}
& (h,h)
\\
&
\multicolumn{3}{c|}{
m_l\geq 1
\;,\;\;
\bar n_k\geq 1
}
\\ \hline
\end{array}
\]
}
\caption{The UV and IR endpoints of the defect flows for a purely
holomorphic perturbation. In each case $\{m_i\}$ and $\{\bar n_j\}$ satisfy
$m_l\geq m_{l{+}1}+2$ and ${\bar n}_k\geq {\bar n}_{k{+}1}+2$.
}
\label{table:1}
\end{table}

\begin{figure}
\begin{centering}
\includegraphics[bb = 0 0 880 540, scale=0.5, type=eps]{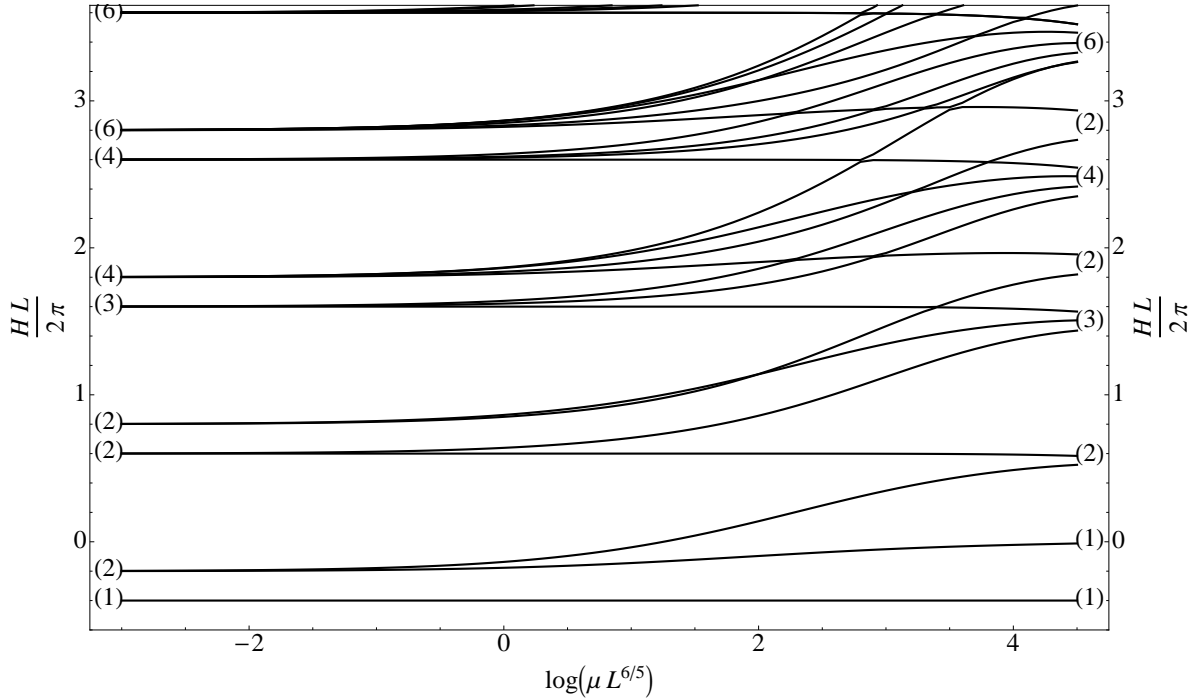}
\par\end{centering}

\caption{Defect flows in the case of a purely holomorphic perturbation
calculated from DTCSA; the subtracted energies $[(E_i - E_0)
L/(2\pi)-2/5]$ are plotted against 
$\log(\mu L^{5/6})$. The flows go from the spectrum of the UV defect
$\cH^{d}$ on the left to the IR defect $\cH^{(0,0)}$ on the
right. The degeneracies of the states are shown in brackets.}

\label{fig:flow05}
\end{figure}

When the energy eigenspaces in question are one-dimensional then the
flows are uniquely defined, as is the case for the flows starting from
the 24 lowest-lying states in $\cH^{d}$. 
Their flows are given in table \ref{tab:flows}.
\begin{table}[htb]
\begin{centering}
\begin{tabular}{|c|c|c|c|}
\hline 
Energy & $\mathcal{H}^{d}$ & $\mathcal{H}^{\left(0,0\right)}$ & Energy\tabularnewline
\hline 
\hline 
$-\frac{2}{5}$ & $\vert D\rangle$ & $\vert h,h\rangle$ & $-\frac{2}{5}$\tabularnewline
\hline 
\multirow{2}{*}{$-\frac{1}{5}$} & $\vert\bar{d}\rangle$ & $\vert0,0\rangle$ & $0$\tabularnewline
\cline{2-4} 
 & $\vert d\rangle$ & $\bar{L}_{-1}\vert h,h\rangle$ & \multirow{2}{*}{$\frac{3}{5}$}\tabularnewline
\cline{1-3} 
\multirow{2}{*}{$\frac{3}{5}$} & $L_{-1}\vert D\rangle$ & $L_{-1}\vert h,h\rangle$ & \tabularnewline
\cline{2-4} 
 & $\bar{L}_{-1}\vert D\rangle$ & $\bar{L}_{-2}\vert h,h\rangle$ & \multirow{2}{*}{$\frac{8}{5}$}\tabularnewline
\cline{1-3} 
\multirow{2}{*}{$\frac{4}{5}$} & $L_{-1}\vert d\rangle$ & $L_{-1}\bar{L}_{-1}\vert h,h\rangle$ & \tabularnewline
\cline{2-4} 
 & $\bar{L}_{-1}\vert\bar{d}\rangle$ & $\bar{L}_{-2}\vert0,0\rangle$ & $2$\tabularnewline
\hline 
\multirow{3}{*}{$\frac{8}{5}$} & $L_{-2}\vert D\rangle$ & $L_{-2}\vert h,h\rangle$ & $\frac{8}{5}$\tabularnewline
\cline{2-4} 
 & $\bar{L}_{-2}\vert D\rangle$ & $\bar{L}_{-3}\vert h,h\rangle$ & \multirow{2}{*}{$\frac{13}{5}$}\tabularnewline
\cline{2-3} 
 & $L_{-1}\bar{L}_{-1}\vert D\rangle$ & $L_{-1}\bar{L}_{-2}\vert h,h\rangle$ & \tabularnewline
\hline 
\multirow{4}{*}{$\frac{9}{5}$} & $L_{-2}\vert\bar{d}\rangle$ & $L_{-2}\vert0,0\rangle$ & $2$\tabularnewline
\cline{2-4} 
 & $L_{-2}\vert d\rangle$ & $L_{-2}\bar{L}_{-1}\vert h,h\rangle$ & $\frac{13}{5}$\tabularnewline
\cline{2-4} 
 & $\bar{L}_{2}\vert\bar{d}\rangle$ & $\bar{L}_{-3}\vert0,0\rangle$ & $3$\tabularnewline
\cline{2-4} 
 & $\bar{L}_{-2}\vert d\rangle$ & $\bar{L}_{-3}\bar{L}_{-1}\vert h,h\rangle$ & $\frac{18}{5}$\tabularnewline
\hline 
\multirow{4}{*}{$\frac{13}{5}$} & $L_{-3}\vert D\rangle$ & $L_{-3}\vert h,h\rangle$ & $\frac{13}{5}$\tabularnewline
\cline{2-4} 
 & $\bar{L}_{-3}\vert D\rangle$ & $\bar{L}_{-4}\vert h,h\rangle$ & \multirow{3}{*}{$\frac{18}{5}$}\tabularnewline
\cline{2-3} 
 & $L_{-2}\bar{L}_{-1}\vert D\rangle$ & $L_{-2}\bar{L}_{-2}\vert h,h\rangle$ & \tabularnewline
\cline{2-3} 
 & $L_{-1}\bar{L}_{-2}\vert D\rangle$ & $L_{-1}\bar{L}_{-3}\vert h,h\rangle$ & \tabularnewline
\hline 
\multirow{6}{*}{$\frac{14}{5}$} & $L_{-3}\vert\bar{d}\rangle$ & $L_{-3}\vert0,0\rangle$ & $3$\tabularnewline
\cline{2-4} 
 & $L_{-3}\vert d\rangle$ & $L_{-3}\bar{L}_{-1}\vert h,h\rangle$ & $\frac{18}{5}$\tabularnewline
\cline{2-4} 
 & $L_{-2}\bar{L}_{-1}\vert\bar{d}\rangle$ & $L_{-2}\bar{L}_{-2}\vert0,0\rangle$ & \multirow{2}{*}{$4$}\tabularnewline
\cline{2-3} 
 & $\bar{L}_{-3}\vert\bar{d}\rangle$ & $\bar{L}_{-4}\vert0,0\rangle$ & \tabularnewline
\cline{2-4} 
 & $\bar{L}_{-3}\vert d\rangle$ & $\bar{L}_{-4}\bar{L}_{-1}\vert h,h\rangle$ & \multirow{2}{*}{$\frac{23}{5}$}\tabularnewline
\cline{2-3} 
 & $L_{-1}\bar{L}_{-2}\vert d\rangle$ & $L_{-1}\bar{L}_{-3}\bar{L}_{-1}\vert h,h\rangle$ & \tabularnewline
\hline 
\end{tabular}
\par\end{centering}
\caption{Defect flows in the case of a purely holomorphic
  perturbation. The flows starting from the lowest-lying 24 states in
  $\cH^{d}$ are uniquely determined by the lattice calculations in
  \cite{BP} given in Table \ref{table:1}.}
\label{tab:flows}
\end{table}
We can identify these flows if we plot the eigenvalues of
${HL}/({2\pi})$ against $\log(L \mu^{5/6})$, as we do in figure \ref{fig:flow05}.

\subsubsection*{Anti-holomorphic perturbation}

Let us introduce a purely anti-holomorphic perturbation of the form
\begin{equation}
  S
= S_{DCFT}-\bar{\mu}\int_{-\infty}^{\infty}dy\, \bar{\varphi}(y)
\;,
\end{equation}
and see how the formulae above change. Clearly only the anti-holomorphic
part is affected now ($\Delta T=0$). An analogous argument and
calculation gives the exact result for the change of the defect condition:
\begin{equation}
  \Delta\bar{T}(y)
= -2\pi\bar{\mu}(1-h)i\partial_{y}\bar{\varphi}(y)
\;.
\end{equation}
 This leads to the conserved energy and momentum in the form
\begin{equation}
  H
= H_{-}+H_{+}+2\pi\bar{\mu}(1-h)\bar{\varphi}
\quad;\qquad 
  P
= P_{-}+P_{+}-2\pi\bar{\mu}(1-h)\bar{\varphi}
\;.
\end{equation}

The anti-holomorphic defect flow can be obtained from the holomorphic one
by a trivial (left-right) replacement.

\subsubsection*{Combined holomorphic and anti-holomorphic perturbations}

We can try to combine holomorphic and anti-holomorphic perturbations
of the form 
\begin{equation}
S=S_{DCFT}-\int_{-\infty}^{\infty}\left(\mu\varphi(y)+\bar{\mu}\bar{\varphi}(y)\right)dy
\end{equation}
 The jump of the chiral half of the energy momentum tensor is given
by

\begin{equation}
  \Delta T(y)
=\lim_{x\to0} \lim_{\epsilon\to 0}
  \left(
  (T_{0}(-x+iy)-T_{0}(x+iy))
  e^{\int_{-\infty}^{\infty}\left(\bar{\mu}\bar{\varphi}(y')+\mu\varphi(y')\right)dy'}
  \right) 
\end{equation}
which has an expansion of the form
\begin{equation}
\Delta T=\mu\mathcal{O}_{1}+\bar{\mu}\mathcal{O}_{\bar{1}}+\mu^{2}\mathcal{O}_{2}
+\bar{\mu}^{2}\mathcal{O}_{\bar{2}}+\mu\bar{\mu}\mathcal{O}_{1\bar{1}}+\dots
\end{equation}
Comparing the dimensions we can write
\begin{equation}
[\mathcal{O}_{n\bar{n}}]=(n+\bar{n})h-n-\bar{n}+2
\end{equation}
clearly we have the previous solutions for $n=1$, $\bar{n}=0$ with
$\mathcal{O}_{1}\propto \partial\varphi$. (Alternatively,  
for $\Delta\bar{T}$ we have $n=0$, $\bar{n}=1$ with
$\mathcal{O}_{\bar{1}}\propto
\bar{\partial}\bar{\varphi}$). Additionally, to these cases we also have the possibility
$n=\bar{n}=1$ 
with either of the two equivalent expressions,
\begin{eqnarray}
\mathcal{O}_{1\bar{1}}(y)
&=&d_+ \varphi(y)\bar\varphi(y) + d_- \bar\varphi(y)\varphi(y) \\
&=&c_{+}\Phi_{+}(y)+c_{-}\Phi_{-}(y)
\end{eqnarray}
In Appendix B we calculate $d_{\pm}, c_{\pm}$ by carefully taking into account
the contribution of $\mathcal{O}_{1}$ to $\Delta T$ at order $\mu\bar{\mu}$.
As a result we obtain 
\begin{equation}
d_\pm = \pm 2\pi i\, h
\;,
\end{equation}
and equivalently
\begin{equation}
c_{\pm}=2\pi ihC_{[\varphi,\bar{\varphi}]}^{\pm}\qquad;\qquad 
C_{[\varphi,\bar{\varphi}]}^{\pm}=C_{\varphi\bar{\varphi}}^{\pm}
-C_{\bar{\varphi}\varphi}^{\pm}
\end{equation}
 Summarising, this means that 
\begin{equation}
  \frac{\Delta T}{2\pi i}
=(1-h) \mu\partial_{y}\varphi
\;+\;
\mu\bar{\mu}hC_{[\varphi,\bar{\varphi}]}^{+}\Delta\Phi\label{eq:DTphiphibar0}
\end{equation}
where $\Delta\Phi=\lim_{x\to+0}(\Phi(-x+iy)-\Phi(x+iy))=\Phi_{-}-\Phi_{+}$.

Some caution is required here:
First note that holomorphic and anti-holomorphic defect fields do not necessarily commute. By conformal invariance their OPE should start with a regular term and if they were bulk fields this would imply that moving one field around the other no monodromy is picked up thus they would commute. Defect fields, however live only on the defects and we do not have the possibility to exchange the two fields without leaving the defect. 
Since the perturbation includes the non-commuting anti-chiral and chiral fields $\bar\varphi$, $\varphi$, the time derivative
$\partial_y\varphi(y)$ taken in the unperturbed theory is not the same as the total time
derivative of the field calculated in the perturbed theory. Instead we have
\begin{equation}
   [\partial_y\varphi]_{tot}
:= \partial_y( e^{-\delta S} \varphi) 
=  e^{-\delta S} 
   \left( \partial_y \varphi + \bar \mu[\varphi,\bar\varphi] \right)
\;.
\end{equation}
Since it is the total time-derivative we are interested in,
we have the final result for the jump in $T$,
\begin{equation}
  \frac{\Delta T}{2\pi i}
= (1-h)\left[\mu\partial_{y}\varphi\right]_{tot}
\;+\;
\mu\bar{\mu}\,C_{[\varphi,\bar{\varphi}]}^{+}\Delta\Phi\label{eq:DTphiphibar}
\;.
\end{equation}

This can be a total time derivative only for chiral perturbations,
i.e. when either $\mu$ or $\bar{\mu}$ vanishes. Similarly we obtain
\begin{equation}
\frac{\Delta\bar{T}}{2\pi i}
= -(1-h)\left[\bar{\mu}\partial_{y}\bar{\varphi}\right]_{tot}
\;+\; 
\mu\bar{\mu}\,C_{[\varphi,\bar{\varphi}]}^{+}\Delta\Phi\label{eq:DbarTphiphibar}
\end{equation}
Clearly in calculating the energy, the jump  $\Delta T-\Delta\bar{T}$
is a total $y$--derivative and so a conserved energy can be defined,
as we expected from time-translation
invariance. This is not true for the momentum, where $\Delta
T+\Delta\bar{T}$ is not a total $y$--derivative. The special form of the 
non-derivative term which does appear however, ($\Delta\Phi$), enables us to cancel it
by introducing an appropriately chosen bulk perturbation.

\section{Massive perturbations}
\label{sec:massp}

We start by analyzing  a purely bulk perturbation without any defect.

\subsubsection*{Pure bulk perturbation}

The perturbed action is given by 
\begin{equation}
S=S_{0}-\lambda\int dudv\,\Phi(u,v)=S_{0}-\lambda\int d^{2}w\,\Phi(w,\bar{w})\label{eq:bulkpert}
\end{equation}
The corresponding change in the conservation law comes from $\bar{\partial}T\neq0$
and can be calculated in a perturbative expansion 
\begin{equation}
\partial_{\bar{z}}T(z)=\partial_{\bar{z}}\left(T_{0}(z)e^{\lambda\int d^{2}w\,\Phi(w,\bar{w})}\right)
=\lambda\mathcal{O}_{1}+\lambda^{2}\mathcal{O}_{2}+\dots
\end{equation}
Dimensional argumentation shows that the only perturbative contribution
comes from the first order term:
\begin{equation}
\partial_{\bar{z}}\left(T_{0}(z)\Phi(w,\bar{w})\right)=\partial_{\bar{z}}
\left(\frac{h\Phi(w,\bar{w})}{(z-w)^{2}}+\frac{\partial_{w}\Phi(w,\bar{w})}{(z-w)}\right)
\end{equation}
We use that 
\begin{equation}
\partial_{\bar{z}}\frac{1}{z-w}=\pi\delta^{(2)}(z-w)\qquad;\qquad
\partial_{\bar{z}}\frac{1}{(z-w)^{2}}=\pi\partial_{w}\delta^{(2)}(z-w)
\end{equation}
and integrate by parts. Assuming fields vanish at infinities we can
drop the surface term and obtain: 
\begin{equation}
\partial_{\bar{z}}T(z)=\pi\lambda(1-h)\partial_{z}\Phi(z,\bar{z})
\equiv\partial_{z}\Theta(z,\bar{z})
\end{equation}
From the dimensional argument we conclude that there are no higher
order terms. We have a similar expression for the anti-holomorphic part
\begin{equation}
\partial_{z}\bar{T}(\bar{z})=\pi\lambda(1-h)\partial_{\bar{z}}
\Phi(z,\bar{z})\equiv\partial_{\bar{z}}\bar{\Theta}(z,\bar{z})
\end{equation}
These conserved currents lead to conserved charges: 
\begin{equation}
H=\int_{-\infty}^{\infty}dx\, (T(z)+\bar{T}(\bar{z})+2\pi\lambda(1-h)
\Phi(z,\bar{z}))\;;\quad P=\int_{-\infty}^{\infty}dx\, (T(z)
-\bar{T}(\bar{z}))\label{eq:HP}
\end{equation}
and so their conservation follows as we have to integrate a total
derivative: 
\begin{equation}
\dot{H}=i\int_{-\infty}^{\infty}dx\, \partial_{x}(T(z)-\bar{T}(\bar{z}))\;;
\quad\dot{P}=i\int_{-\infty}^{\infty}dx\, \partial_{x}(T(z)+\bar{T}(\bar{z})
-2\pi\lambda(1-h)\Phi(z,\bar{z}))
\end{equation}

If we introduce the defect, then the local conservation laws are not
changed but we have to be careful with the surface terms at the
defect.
As before, we will cut off all perturbative integrals at a distance
$\epsilon$ and we will take $\epsilon\to 0$ before any other limits. 

Using this convention, we find in appendix \ref{app:details} that the
bulk perturbation introduces jumps in the energy momentum tensor of
the form 
\begin{equation}
  \Delta T(y)
= -\lambda\pi h\,\Delta\Phi(y)
\;,\;\;
  \Delta \bar T(y)
=- \lambda\pi h\,\Delta\Phi(y)
\end{equation}
Defining $H_{\pm}$
and $P_{\pm}$ by splitting the integrals in eq. (\ref{eq:HP}) as
we did in eq. (\ref{eq:Hp+Hm},\ref{eq:Pp+Pm}): 
\begin{equation}
H=H_{-}+H_{+}\quad;\qquad P=P_{-}+P_{+}
\end{equation}
we can easily see that 
\begin{equation}
  \partial_{y}H
= i\left[\Delta T-\Delta\bar{T}\right]
= 0
\;,\;\;\;\;
  \partial_{y}P
= i\left[\Delta T+\Delta\bar{T}-2\pi\lambda(1-h)\Delta\Phi\right]
= - 2\pi i\,\lambda\,\Delta\Phi
\neq0
\end{equation}
Clearly the defect perturbation, without any defect field is not integrable. As the form
of $\partial_{y}P$ is the same as the contribution of the combined
holomorphic and anti-holomorphic defect perturbation:
eq. (\ref{eq:DTphiphibar},\ref{eq:DbarTphiphibar}) 
by properly synchronizing their coefficients we can ensure integrability.

\subsubsection*{Combined bulk and defect perturbation}

Now we introduce simultaneously the bulk perturbation and the chiral and anti-chiral
defect perturbations:

\begin{equation}
  S
= S_{DCFT}
- \lambda\int d^{2}w\,\Phi(w,\bar{w})
- \int_{-\infty}^{\infty}\left(\mu\varphi(y)+\bar{\mu}\bar{\varphi}(y)\right)dy
\end{equation}
From appendix B we see that the jumps in $T$ and $\bar T$ in the case
of the combined perturbation are
\begin{eqnarray}
  \Delta T(y)
&=& 2\pi i(1-h)\mu [\partial_y\varphi]_{tot}
 + (2\pi i \mu\bar\mu C^+_{[\varphi,\bar\varphi]} - \lambda\pi
 h)\Delta \Phi
\;,
\\
  \Delta \bar T(y)
&=& -2\pi i(1-h)\bar\mu [\partial_y\bar\varphi]_{tot}
 + (2\pi i \mu\bar\mu C^+_{[\varphi,\bar\varphi]} - \lambda\pi
 h)\Delta \Phi
\;.
\end{eqnarray}
Using the bulk conservation laws, we find
\begin{eqnarray}
  \partial_y(H_- + H_+)
&=& i(\Delta T - \Delta\bar T)
\nonumber\\
&=& -2\pi(1-h)[\partial_y(\mu\varphi+\bar\mu\bar\varphi)]_{tot}
\;,
\end{eqnarray}
is always a total $y$--derivative and hence the total energy $H$
defined as
\begin{equation}
H = H_- + H_+ + 2\pi(1-h)(\mu\varphi+\bar\mu\bar\varphi)
\;,
\end{equation}
is always conserved.

We also find
\begin{eqnarray}
  \partial_y(P_- + P_+)
&=& i(\Delta T + \Delta\bar T - 2\pi\lambda(1-h)\Delta\Phi)
\nonumber\\
&=& -2\pi(1-h)[\partial_y(\mu\varphi-\bar\mu\bar\varphi)]_{tot}
- (4\pi\mu\bar\mu C^+_{[\varphi,\bar\varphi]} +
2i\lambda\pi)\Delta\Phi
\end{eqnarray}
is a total derivative if $\lambda=2i\mu\bar\mu C^+_{[\varphi,\bar\varphi]}$.
Hence, we can define a total momentum
\begin{equation}
P  =  P_{-}+P_{+}+2\pi(1-h)(\mu\varphi-\bar\mu\bar\varphi)
\;,
\end{equation}
which is conserved exactly when
\begin{equation}
  \lambda 
= 2 i \mu\bar\mu C^+_{[\varphi,\bar\varphi]}
= (0.826608\ldots)\,\mu\,\bar\mu
\;.
\label{eq:lmumu}
\end{equation}
This agrees with the result in \cite{Runkel:2010ym} where the problem
was analysed in the opposite channel.
We can conclude that the perturbation is integrable only if this constraint
is satisfied. As $\lambda$ defines the mass scale, the space of integrable
defect perturbations has one physical parameter. Observe also that
we cannot switch off the defect perturbations completely if we insist
on keeping integrability.

\section{Defect TCSA}
\label{sec:dtcsa}

In this section we review the TCSA method for periodic boundary
conditions and generalize it for the defect case. 

The theory is defined on the cylinder of circumference $L$. The periodic
Hilbert space takes the form 
\begin{equation}
\mathcal{H}=V_{0}\otimes\bar{V}_{0}+V_{1}\otimes\bar{V}_{1}
\end{equation}
 where the unperturbed Hamiltonian acts as 
\begin{equation}
H_{0}=\frac{2\pi}{L} \left( L_{0}+\bar{L}_{0}-\frac{c}{12} \right) \label{eq:H0}
\end{equation}
 The perturbation which defines the scaling Lee-Yang model on the
cylinder is given by 
\begin{equation}
  H = H_{0}-\lambda\int_{0}^{L}\Phi(x,0)dx
\;.
\end{equation}
 Mapping the cylinder onto the plane, ($\zeta=x+iy\to z=e^{-i\frac{2\pi}{L}\zeta}=re^{i\theta}$,
$\bar{z}=e^{i\frac{2\pi}{L}\bar{\zeta}}=re^{-i\theta}$), we find the
Hamiltonian is given by
\begin{equation}
  H
= \frac{2\pi}{L}
  \left[\;
   L_{0}+\bar{L}_{0}+\frac{11}{30}
+ \lambda\left(\frac{L}{2\pi}\right)^{2+\frac{2}{5}}
  \!\int_{0}^{-2\pi} \!\!\!d\theta\,\Phi(e^{i\theta},e^{-i\theta})
  \;\right]
\;.
\end{equation}
The rotation operator on the plane $L_{0}-\bar{L}_{0}$ corresponds
to the momentum operator on the cylinder $P=\frac{2\pi}{L}(L_{0}-\bar{L}_{0})$.
As a result the $\theta$-dependence of the matrix elements of the
perturbing operator can be easily evaluated and the integral gives
momentum conservation: 
\begin{equation}
  \int_{0}^{-2\pi}\langle j\vert\Phi(e^{i\theta},e^{-i\theta})\vec k d\theta
= \int_{0}^{-2\pi}e^{i\theta(h_{k}-\bar{h}_{k}-h_{j}+\bar{h}_{j})}
  d\theta\,\Phi_{jk}
= -\Phi_{jk}2\pi\delta_{P_{k}-P_{j}}
\end{equation}
where $\Phi_{jk}=\langle j\vert\Phi(1,1)\vert k\rangle$ and we used that 
$h_k-\bar{h}_k \in \mathbb{Z} $. Introducing
the inner product matrix $G_{ij}=\langle i\vert j\rangle$ and the
mass gap relation (\ref{eq:ml}) the dimensionless Hamiltonian can be written as 
\begin{equation}
  \frac{H}{m}
= \frac{2\pi}{mL}
   \left[\;
   L_{0}+\bar{L}_{0}+\frac{11}{30}
  -\left(\frac{mL}{2\pi\kappa}\right)^{\frac{12}{5}}
   (2\pi)\, G^{-1}\Phi\;
   \right]
\end{equation}

\subsection*{Pure defect perturbation}

The defect conformal Hilbert space contains the modules 
\begin{equation}
\mathcal{H}^{d}=V_{1}\otimes \bar{V}_{0}+V_{0}\otimes \bar{V}_{1}
+V_{1}\otimes \bar{V}_{1}
\end{equation}
and the unperturbed Hamiltonian is given by eq. (\ref{eq:H0}). 

We start by analyzing a chiral defect perturbation of the form
\begin{equation}
H=H_{0}-\mu\varphi(y=0)
\end{equation}
We map the cylinder to the conformal plane ($\zeta=x+iy\to z=e^{-i\frac{2\pi}{L}\zeta}$,
such that the defect $x=0$ will fill the real positive line: $z=e^{\frac{2\pi}{L}y}$). As
the defect field is chiral it will acquire an additional phase, $\rho=e^{i\pi/10}$. 
To distinguish between the cases when the defect is located on the imaginary 
$\varphi (iy) $ or on the real $\varphi(x)$ line we introduce another coupling 
$\hat \mu = \mu \rho$, such that 
\begin{equation}
\mu \varphi(i x)= \hat \mu \varphi( x) \qquad ;\quad \hat \mu = \mu \rho \quad ;\quad 
\rho =e^{i\frac{\pi}{10}} 
\end{equation}
With this coupling the Hamiltonian on the plane is 
\begin{equation}
H=\frac{2\pi}{L}\left[L_{0}+\bar{L}_{0}-\frac{c}{12}-
\hat \mu\left(\frac{L}{2\pi}\right)^{1+\frac{1}{5}}\varphi(1)\right]
\end{equation}
For numerical evaluation we will need the various matrix elements
of $\varphi$, which are evaluated in Appendix A: 
\begin{equation}
C_{\varphi D}^{D}=C_{\varphi\varphi}^{\varphi}=
\alpha\beta^{-1}\quad;\quad C_{\varphi d}^{d}=\,\alpha\beta\quad;
\quad C_{\varphi D}^{\bar{d}}=-C_{\varphi\bar{d}}^{D}=1
\end{equation}
where 
\begin{equation}
\alpha=\sqrt{\frac{\Gamma(\frac{1}{5})\Gamma(\frac{6}{5})}{\Gamma(\frac{3}{5})\Gamma(\frac{4}{5})}}
\quad;\qquad\beta=\sqrt{\frac{2}{1+\sqrt{5}}}
\end{equation}

\subsection*{Combined bulk and defect perturbation}

Now we perturb the conformal defect theory simultaneously in the bulk
and at the defect
\begin{equation}
H=H_{0}-\lambda\int_{0}^{L}\Phi(x,0)dx
-\mu\varphi(0)- \bar{\mu}\bar{\varphi}(0)
\end{equation}
Mapping the system onto the plane
\begin{equation}
  H
= \frac{2\pi}{L}
  \left[\;
  L_{0}+\bar{L}_{0}+\frac{11}{30}
- \left(\frac{L}{2\pi}\right)^{\!1+\frac{1}{5}}\left(
\hat \mu\varphi(1)+\hat{\bar{\mu}}\bar{\varphi}(1)\right)
+ \lambda\left(\frac{L}{2\pi}\right)^{\!2+\frac{2}{5}}
  \!\int_{0}^{-2\pi}d\theta\,\Phi(e^{i\theta},e^{-i\theta})
  \;\right]
  \label{TCSAH}
\end{equation}
where $ \hat{\bar{\mu}}= \bar \mu  \rho^{-1} $. 
Using the rotation symmetry we can perform the integrals
\begin{equation}
\rho_{jk}:=\int_{0}^{-2\pi}e^{i\theta S_{jk}}d\theta=\begin{cases}
\begin{array}{c}
-2\pi\\
-2e^{-i\pi S_{jk}}\frac{\sin\pi S_{jk}}{S_{jk}}
\end{array} & \begin{array}{c}
\mbox{ if }S_{jk}=0\\
\mbox{otherwise}
\end{array}\end{cases} 
\end{equation}
where the difference of the spins
\begin{equation}
S_{jk}:= h_{j}-\bar{h}_{j}-h_{k}+\bar{h}_{k}
\end{equation}
is usually not an integer. 
The matrix form of the dimensionless Hamiltonian is simply 
\begin{equation}
  \frac{H}{m}
= \frac{2\pi}{mL}
  \left[L_{0}+\bar{L}_{0}+\frac{11}{30}
     -\left(\frac{L}{2\pi}\right)^{\frac{6}{5}}G^{-1}
      \left(\hat \mu\varphi+\hat{\bar{\mu}}\bar{\varphi}\right)
     +\left(\frac{mL}{2\pi\kappa}\right)^{\frac{12}{5}}\, G^{-1}\,\Phi_{\rho}
  \right]
\end{equation}
where $\langle j\vert\Phi_{\rho}\vec k =\Phi_{jk}\rho_{jk}$. 

The relevant structure constants are
\begin{equation}
\ensuremath{C_{\varphi\bar{\varphi}}^{\Phi_{-}}=C_{\bar{\varphi}\varphi}^{\Phi_{+}}}=
\frac{\beta^{-1}}{1+\eta^{-1}}\qquad;\qquad C_{\varphi\bar{\varphi}}^{\Phi_{+}}=
C_{\bar{\varphi}\varphi}^{\Phi_{-}}=\frac{\beta^{-1}}{1+\eta}
\end{equation}
\begin{equation}
C_{\Phi_{-}d}^{\bar{d}}=-\eta^{-2}\beta^{-1}\quad;
\quad C_{\Phi_{-}\bar{d}}^{d}= -\eta^2\beta^{-1}\quad; \quad
C_{\Phi_{-}d}^{D}=-\eta^{-1}\alpha\quad; \quad C_{\Phi_{-}D}^{\bar{d}}= \eta^{-1} \alpha 
\end{equation}
\begin{equation}
C_{\Phi_{-}\bar{d}}^{D}= -\eta \alpha \quad ;\quad C_{\Phi_{-}D}^{d}=\eta \alpha  
\quad;\quad C_{\Phi_{-}D}^{D}=\alpha^{2}\beta^{-1}
\end{equation}

\subsection*{Integrability in DTCSA}

It is interesting to analyze the integrability of the model by demanding
the commutation of energy and momentum $[H,P]=0.$ The momentum in
the TCSA scheme is given by 
\begin{equation}
P=\frac{2\pi}{L}\left[L_{0}-\overline{L}_{0}-\left(\frac{L}{2\pi}\right)^{\frac{6}{5}}
\hat \mu\varphi(1)+\left(\frac{L}{2\pi}\right)^{\frac{6}{5}}\hat{\bar{\mu}}
\bar{\varphi}(1)\right]
\end{equation}
while the energy by (\ref{TCSAH}). 
We perform the analysis for $L=2\pi$. 

The term
$\left[L_{0}+\overline{L}_{0},\ \hat\mu\varphi(1)-\hat{\bar{\mu}}\bar{\varphi}(1)\right]$ 
cancels against
$\left[\hat\mu\varphi(1)
 + \hat{\bar{\mu}}\bar{\varphi}(1),\ L_{0}-\overline{L}_{0}\right].$ 
Using the identity
$\left[L_{0},\ \Phi\left(z,\bar{z}\right)\right]
=h\Phi(z,\bar{z})+z\partial_{z}\Phi(z,\bar{z})$,   
its anti-holomorphic part together with
$z\partial_{z}-\bar{z}\partial_{\bar{z}}=-i\partial_{\theta,}$ 
we can write
\begin{eqnarray}
  \lambda\int_{0}^{-2\pi}
  \left[\Phi\left(e^{i\theta},e^{-i\theta}\right),\ 
  L_{0}-\overline{L}_{0}\right]d\theta 
& = & 
  \lambda
  i\int_{0}^{-2\pi}\partial_{\theta}\Phi(e^{i\theta},e^{-i\theta})d\theta
\nonumber \\
& = & 
  -i\lambda\left(\Phi_{+}(1)-\Phi_{-}(1)\right)
\;.
\label{eq:07}
\end{eqnarray}
This term has to cancel against
$ 2\hat\mu\hat{\bar{\mu}}\left[\bar{\varphi}(1),\ \varphi(1)\right]
 = -2 \hat\mu\hat{\bar{\mu}} C^+_{[\varphi,\bar\varphi]} (\Phi_+ - \Phi_-)$ 
which leads to
\begin{equation}
 \hat\mu \hat{\bar{\mu}}= \mu\bar{\mu}
= \lambda\frac{1}{2i C^+_{[\varphi,\bar\varphi]}}
\;,\;\;\;\;
  C^+_{[\varphi,\bar\varphi]}
= C_{\varphi\bar{\varphi}}^{\Phi_+}-C_{\bar\varphi\varphi}^{\Phi_+}
= \beta^{-1}\frac{1-\eta}{1+\eta}
= -i \,0.413304 \ldots 
\end{equation}
This result is the same as we calculated before. 

Finally we analyze
the term 
\begin{equation}
\int_{0}^{-2\pi}\left[\Phi\left(e^{i\theta},e^{-i\theta}\right),\ 
\hat\mu\varphi(1)-\hat{\bar{\mu}}\bar{\varphi}(1)\right]d\theta.
\label{eq:lastterm}
\end{equation}
Let's denote $\psi(x)=\hat\mu\varphi(x)-\hat{\bar{\mu}}\bar{\varphi}(x)$. In
taking the products of operators, they have to be radially ordered,
therefore in the commutator the contour of the integration is deformed
by $\epsilon$: in the $\Phi\psi$ term the radius of the integration
is $1+\epsilon$, while in the $\psi\Phi$ term the radius is $1-\epsilon$.
Then, the contour of the integration can be transformed: one integral
from $1+\epsilon$ to $1-\epsilon$ on the upper side of the defect
plus one integral from $1-\epsilon$ to $1+\epsilon$ on the lower
side of the defect. The limit of $\Phi$ on the defect from above
is $\Phi_{-}$, and the limit from below is $\Phi_{+}$. We can use
the OPEs to calculate these integrals. The OPEs of $\Phi_{+}$ and
$\Phi_{-}$ with $\varphi$ and $\bar{\varphi}$ are regular in $\epsilon$,
and we can perform the integration. After the integration we get only
positive power terms in $\epsilon$ which are vanishing in the $\epsilon\rightarrow0$
limit, and so (\ref{eq:lastterm}) is zero.

\section{Scattering description of defects in the scaling Lee-Yang model}
\label{sec:6}

We summarise here the results of 
\cite{Bajnok:2007jg} on the integrable description of defects in the
Lee-Yang model and give the UV-IR correspondence relating the
parameters in the integrable and perturbed DCFT descriptions. 

The scaling Lee-Yang model has a single massive particle with
mass
\begin{equation}
  m
= \kappa\,\lambda^{\frac{5}{12}}
\;,\;\;\;\;
  \kappa
= \frac{2^{19/12}\sqrt{\pi}}{5^{5/16}}\,
  \frac{(\Gamma(3/5)\Gamma(4/5))^{5/12}}{\Gamma(2/3)\Gamma(5/6)}\,
= 2.642944
\ldots
\;,
\label{eq:ml}
\end{equation}
and two-particle $S$--matrix
\begin{equation}
S(\theta) = -\left(\frac 13\right)\left( \frac 23 \right)
\;,\;\;\;\;
  (x) 
= \frac{\sinh(\frac\theta 2 + \frac{i\pi x}2)} 
       {\sinh(\frac\theta 2 - \frac{i\pi x}2)} 
\;.
\end{equation}
An integrable defect is described by two transmission factors,
$T_-(\theta)$ for a particle crossing from left to right with rapidity
$\theta>0$ and $T_+(-\theta)$ for a particle crossing from right to left
with rapidity $\theta<0$.
The authors of \cite{Bajnok:2007jg} proposed the following one-parameter family of
solutions to the fusion, crossing and unitarity relations:
\begin{equation}
  T_{-}
= \left[b+1\right]\left[b-1\right]
= S \Bigl ( \theta - \frac{i \pi (3{-}b)}6\Bigr )
\;,\;\;
  T_{+}
= \left[5-b\right]\left[-5-b\right]
= S \Bigl ( \theta + \frac{i \pi (3{-}b)}6\Bigr )
\;,\;\;
\end{equation}
where
\begin{equation}
  \left[x\right]
= i\frac{\sinh\left(\frac{\theta}{2}+i\frac{\pi x}{12}\right)}
        {\sinh\left(\frac{\theta}{2}+i\frac{\pi
            x}{12}-i\frac{\pi}{2}\right)}
\;.
\end{equation}
Thus it can be seen that the defect is equivalent, for scattering
purposes, to a particle with rapidity $i\pi(3-b)/6$, and the
transmission factor is a pure phase for $b=\mp3 + i\alpha$.

 According to
\cite{Bajnok:2007jg}
the bulk energy-density and the infinite volume defect energy are
\begin{equation}
  \epsilon_{bulk}
= -\frac{1}{4\sqrt{3}}m^{2}
\qquad;\quad
  \epsilon_{Def}
= m\sin\left(\frac{b\pi}{6}\right)
\;,
\label{eq:en}
\end{equation}
and the finite size corrections for the ground state energy are also given,
in first order, by the L\"uscher correction term which is
\begin{equation}
  E_{0}\left(L\right)
= - m\int_{-\infty}^{\infty}
   \frac{d\theta}{2\pi}\cosh\left(\theta\right)
   T_+ \left( \frac{ i\pi }{2}-\theta \right)
   e^{-mL\cosh\left(\theta\right)}
 + O\left(e^{-2mL}\right)
\;.
\label{eq:lusch}
\end{equation}
The Defect Thermodynamic Bethe Ansatz (DTBA) equations were also derived. The pseudo energy is given as the solution of the integral equation
\begin{equation}
 \varepsilon \left( \theta \right) = m L \cosh \theta - \int_{-\infty}^{\infty} \frac{d \theta'}{2 \pi} \phi \left( \theta-\theta' \right) \log \left( 1+ T_{+} \left(\frac{i \pi}{2}-\theta' \right) e^{- \varepsilon\left( \theta' \right) }\right)
 \label{eq:DTBAepsilon}
\end{equation}
where $\phi (\theta) = -i \frac{d}{d\theta} \log S(\theta)$. The ground state energy is expressed via the pseudo energy as
\begin{equation}
E_{0} \left( L \right) = -m \int_{-\infty}^{\infty} \frac{d \theta}{2 \pi} \cosh \left(\theta\right) \log \left( 1 + T_{+} \left( \frac{i \pi}{2} -\theta \right) e^{-\varepsilon \left( \theta \right)}\right)  
\label{eq:DTBAenergy}
\end{equation}
The DTBA equations are reliable at least for such values of the defect parameter $b$ when the transmission factor $T_{+}$ is a pure phase, i. e. for $b=-3+i\alpha$ with real $\alpha$.

There are several ways to derive the UV-IR correspondence. One is by comparing
the action of the defect on the identity boundary condition with the
perturbed boundary condition $\Phi(b)$. As the defect approaches the
boundary, the two defect fields $\varphi(x)$ and $\bar\varphi(x)$ both
have the same limit, the relevant boundary field $\phi(x)$\footnote{This is true when the defect and the boundary are both oriented along the real axis otherwise the fields acquire relative phases}, so that the
defect perturbation with parameters $(\hat\mu,\hat{\bar{\mu}})$ becomes the
boundary perturbation with parameter $h=\hat \mu+\hat{\bar{\mu}}$, 
$\hat \mu\varphi + \hat{\bar{\mu}}\bar\varphi \to (\hat \mu+\hat{\bar{\mu}})\phi$. 
The boundary UV-IR relation is \cite{Dorey:1997yg}:
\begin{eqnarray}
&&  h 
= |h_c|\cos((b+3)\pi/5) m^{6/5}
= \frac{|h_c|m^{6/5}}2\,e^{(b+3)\pi i/5}
\;+\;
  \frac{|h_c|m^{6/5}}2\,e^{-(b+3)\pi i/5}
\;,\;\;\;\;
\\ 
&&
h_c 
= -\frac{\pi^{3/5}2^{4/5} 5^{1/4}\sin \frac{2\pi}5}
        {(\Gamma(\frac 35)\Gamma(\frac 45))^{1/2}}
   \left(\frac{\Gamma(\frac 23)}{\Gamma(\frac 16)}\right)^{6/5}
= -0.685289\ldots
\;.
\end{eqnarray}
The natural identification is
\begin{equation}
    \hat \mu =   \frac{|h_c|m^{6/5}}2\,e^{\pm i(b+3)\pi/5}
\;,\;\;
\hat{\bar{\mu}} =   \frac{|h_c|m^{6/5}}2\,e^{\mp i(b+3)\pi/5}
\;.
\end{equation}
It is easy to check that
\begin{equation}
  \lambda 
=  \left(\frac{4}{h_c^2\,\kappa^{12/5}}\right) \hat \mu\,\hat{\bar{\mu}}
\;,
\end{equation}
agrees with the integrability condition (\ref{eq:lmumu}).

The ambiguity in the exponent can be checked in several ways: one is
by considering the behaviour of the $T$--matrices for $b=-3+i\alpha$ in
the two limits $\alpha\to\pm\infty$.
In both these limits, $T_\pm(\theta)\to 1$ for any $\theta$, but not in a uniform
fashion. 

In the limit $\alpha\to+\infty$, $T_+(\theta)$ does tend to 1
uniformly, but $T_-(\theta)$ changes rapidly around
$\theta\sim\alpha/2$ indicating that the defect has no effect on
left-moving modes but a large effect on right moving modes in the far
UV; this behaviour corresponds to - in our convention of the complex coordinates - a purely holomorphic perturbation of the
topological defect with $\hat \mu\to \infty$ and $\hat{\bar{\mu}}\to 0$ in this limit.

Conversely, 
in the limit $\alpha\to-\infty$, $T_-(\theta)$  tends to 1
uniformly, but $T_+(\theta)$ changes rapidly around
$\theta\sim\alpha/2$ indicating that the defect has no effect on
right-moving modes but a large effect on left-moving modes in the far
UV, corresponding to a purely anti-holomorphic
(affecting the left-moving modes only) perturbation of the
topological defect so that $\hat{\bar{\mu}}\to \infty$ and $\hat\mu\to 0$ in this limit.

Using this, we see that the correct identification is
\begin{eqnarray}
&&
  \hat\mu 
= \frac{|h_c|m^{6/5}}2\,e^{ \alpha\pi/5}
= \frac{|h_c|m^{6/5}}2\,e^{  -i(b+3)\pi/5}
= -\frac{|h_c|m^{6/5}}2\,e^{  -i(b-2)\pi/5}
\;,\;\;
\\
&&
  \hat{\bar{\mu}} 
= \frac{|h_c|m^{6/5}}2\,e^{- \alpha\pi/5}
= \frac{|h_c|m^{6/5}}2\,e^{ i(b+3)\pi/5}
= -\frac{|h_c|m^{6/5}}2\,e^{ i(b-2)\pi/5}
\;.
\end{eqnarray}

\section{Numerical results}
\label{sec:numer}

We analyzed the numerical spectrum for four choices of the $b$ defect
parameter, namely $b=-3+2i, b=-3, b=0.5$ and $b=1.8$; the first two
were chosen to correspond to the transmission matrix being a phase;
the second two have non-phase scattering but have bound states. We
considered various aspects of the spectra, as follows.

First, we analyzed the ground states.
We numerically solved the ground state energy Lüscher correction equation
for different values of the $b$ defect parameter, and plotted together
with the TCSA ground states. For $b=-3$ and $b=-3+2i$ these lines
fit the TCSA points within one percent for volumes $mL>1$, but in
the two other case, they fit only for $mL>5$, showing that the higher
order finite size corrections should be taken into account. 

\begin{figure}
\begin{centering}
\includegraphics[bb = 0 0 880 561, scale=0.5, type=eps]{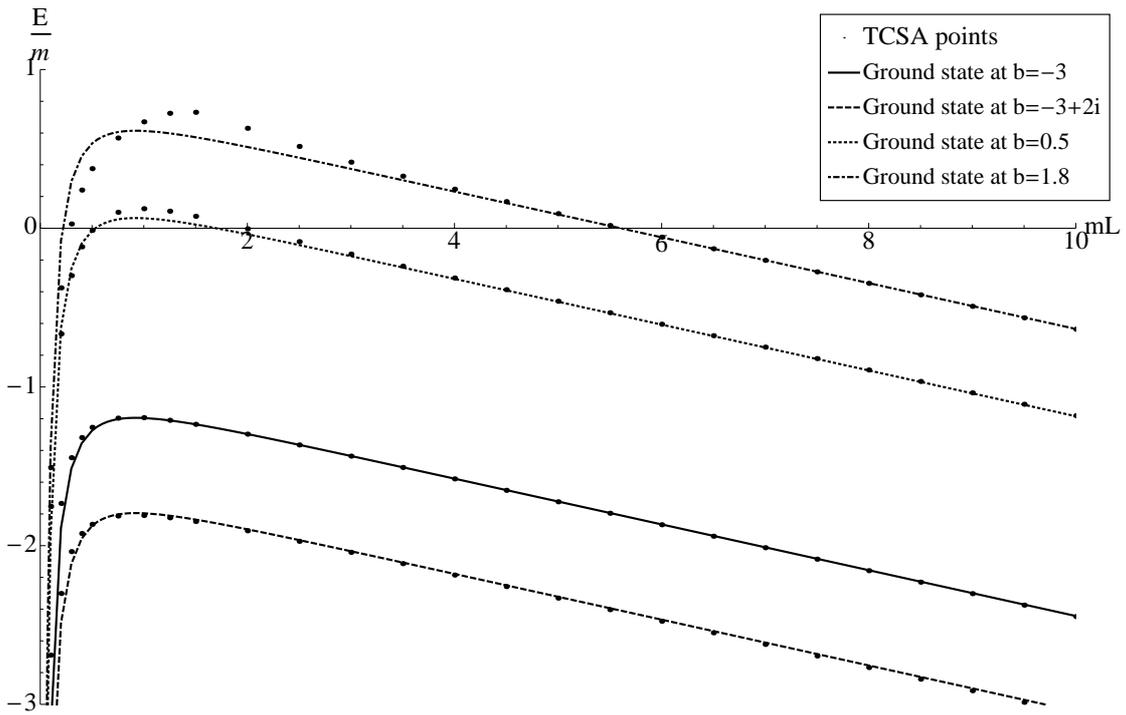}
\par\end{centering}
\caption{Ground state energies with Lüscher corrections and the TCSA ground
state points}
\label{Groundstate}
\end{figure}

\begin{figure}
\begin{centering}
\includegraphics[bb = 0 0 880 561, scale=0.5, type=eps]{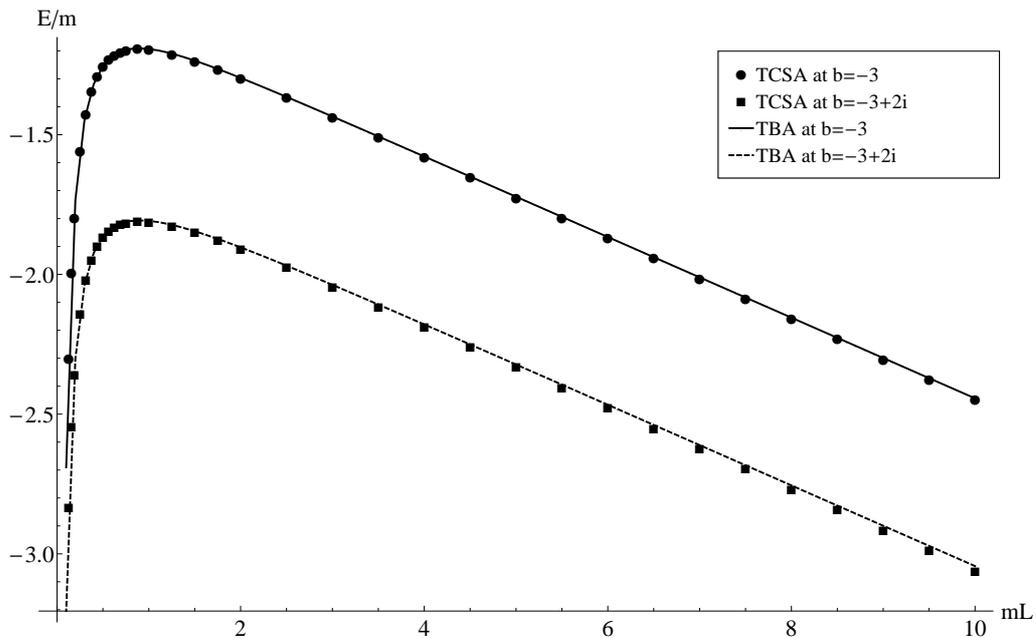}
\par\end{centering}
\caption{Ground state energies from the defect TBA and the TCSA ground
state points}
\label{GroundstateTBA}
\end{figure}

For $b=-3$ and $b=-3+2i$ we solved the equations (\ref{eq:DTBAepsilon}, \ref{eq:DTBAenergy}) iteratively and plot against the TCSA spectrum. This is shown on figure \ref{GroundstateTBA}.

If we choose $b=-3$ the TCSA spectra remains real, and on the scattering
theory side, the transmission factor is just a phase for real rapidity
and has no poles
in the physical strip,  so that we do not expect any defect
bound-states.

\begin{figure}
\begin{centering}
\includegraphics[bb = 0 0 800 540, scale=0.5, type=eps]{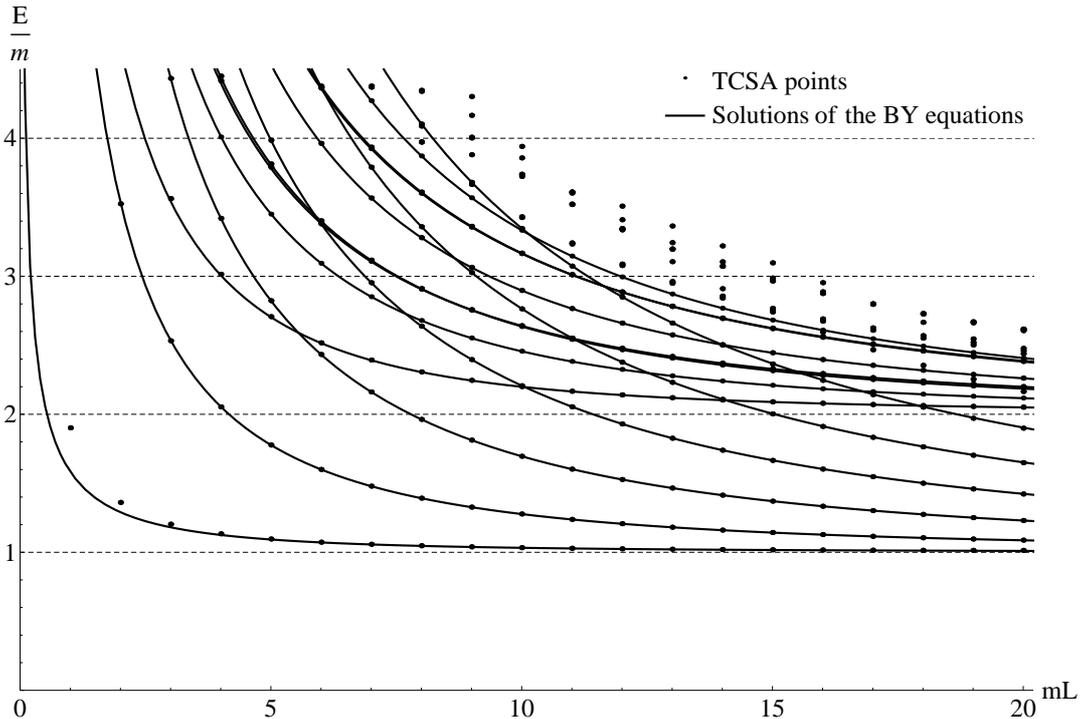}
\par\end{centering}

\caption{The TCSA spectra together with the solutions of the Bethe Yang equations
at $b=-3$.}
\label{BYlines}
\end{figure}

The DTBA equations can be generalised to include the excited states
but instead we used a simpler approximate method which is nevertheless
accurate for volumes that are not too small. 
In finite (but not too small) volumes the solutions of the Bethe-Yang equations,
\begin{equation}
e^{iL\sinh\theta_{i}}T_{-}\left(\theta_{i}\right)\prod_{j\neq i}
S\left(\theta_{i}-\theta_{j}\right)=1\qquad;\quad i=1,\dots,n
\end{equation}
give a good approximation to the rapidities of the $n$-particle
state. From these one can easily calculate the energy of
the $n$-particle state. As at $b=-3$ the transmission factor is
just a phase, we can take the logarithm of these equations, which
become a system of real algebraic equation with $n$ integer parameters
called the Bethe-Yang quantum numbers. We solved these
equations numerically in the case of one and two particles, for the
smallest Bethe-Yang 
quantum numbers, and plotted the resulting energies together with the
modified TCSA spectra which can be obtained from the original spectra
by subtracting the values of the ground state -- see Figure \ref{BYlines}.

We should notice that at $b=-3$, the two transmission factors $T_{-}$
and $T_{+}$ are identical, so that we have exact parity-symmetry
in this case: the right-moving particle has the same energy as the
left-moving. This can be seen in the numerical spectra: all one-particle
Bethe-Yang lines and the corresponding TCSA points have multiplicity
two. 
Among the two-particle Bethe-Yang lines, and the corresponding
TCSA points, the only lines with multiplicity one are those that
correspond to parity-invariant sets of momenta; the others have
multiplicity two.  

\begin{figure}
\begin{centering}
\includegraphics[bb = 0 0 800 540, scale=0.5, type=eps]{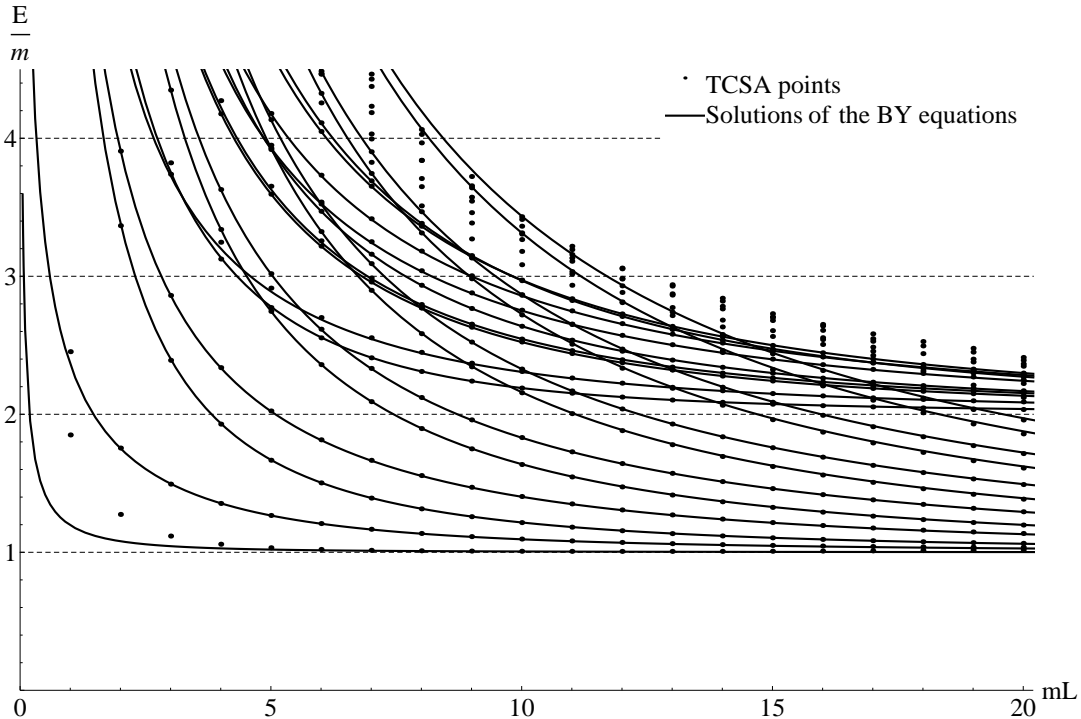}
\par\end{centering}

\caption{The TCSA spectra together with the solutions of the Bethe-Yang equations
at $b=-3+2i$}
\label{fig:extrafig}

\end{figure}

If we choose $b=-3+\alpha i$, with real $\alpha$, the transmission
factor remains a phase, and the TCSA spectrum is real. The Bethe-Yang
equations become a system of real algebraic equations which can be
solved numerically for different quantum numbers. We solved them at
the $b=-3+2i$, and plotted the resulting energies in Figure
\ref{fig:extrafig}
for the smallest
quantum numbers in the case 
of one and two particles, together with the TCSA points. For $mL>5$
every Bethe-Yang line fits the TCSA points. For smaller volumes, 
due to finite size corrections, there is a mismatch, mainly for the
lowest energy lines.

However it is not true any more that the two transmission factors
are identical, the parity-symmetry is broken. Due to this
fact, the two-fold degeneracy of the states which was valid for $b=-3$
is broken; thus for $b=-3+i\alpha$ every Bethe-Yang line and
the corresponding TCSA points have multiplicity one.

If we choose $\Re\mathrm{e}\left(b\right)\neq-3$, the transmission
factors are not just phases any more, and the TCSA spectrum becomes
complex as well. If we take the logarithm of the Bethe-Yang equations,
they become a system of algebraic equations of complex quantities,
each equation holds for both the real and the imaginary part. 
In the following we plot only the real part as it contains the real 
vacuum and the boundary bound-states.

According to 
\cite{Bajnok:2007jg} in different domains of the
parameter $b$, the transmission factor $T_{+}$ has poles, and we
have defect bound states in infinite volume. In the domain $b\in\left[-1,1\right]$
we expect one defect bound state, and if $b\in\left[1,2\right]$ we
expect two. The infinite volume defect energies are given as
\begin{equation}
\epsilon_{1}=m\cos\left(\frac{\pi}{6}\left(b+1\right)\right)\qquad;\quad\epsilon_{2}
=m\cos\left(\frac{\pi}{6}\left(b-1\right)\right)
\end{equation}

One of the values we have chosen, $b=0.5$, corresponds to a system
where we expect a single defect bound state in infinite volume.
In finite volume, a defect bound state corresponds to a solution of
the Bethe-Yang equations with purely imaginary rapidity. For $mL>5.5$,
we in fact find two solutions, one which asymptotically approaches the
defect bound state energy in infinite volume, and one which approaches
a free massive particle state. For smaller volumes, these converge and
meet at $mL \sim 5.5$ and for smaller volumes there are no purely
imaginary solutions to the Bethe-Yang equations and indeed the DTCSA
has complex spectrum and is consistent with complex rapidity solutions.

We can identify the one and two particle states as the solutions of
the Bethe-Yang equations for complex rapidities. In case of the two-particle
Bethe-Yang equations there are two kinds of such solutions: one where
none of the rapidities are purely imaginary, and one where one of
these rapidities is purely imaginary. This latter case corresponds
in infinite volume to one particle scattering on the excited defect
i.e. a defect with one particle bound on it. The two particle Bethe-Yang
equations with one purely imaginary rapidity have solutions only for
larger volumes, $mL>7$, but we have to remember that the Bethe-Yang
equations are not exact, 
in small volumes the vacuum polarisation effects become considerable,
and we can trust these solutions only for these larger volumes.

We plotted the TCSA spectra together with the solutions of the Bethe-Yang
equations for small quantum numbers: The one-particle solutions with
purely imaginary rapidities for $mL>5$, the one-particle solutions
with non-purely imaginary rapidities, the two-particle solutions with one
imaginary rapidity for $mL>7$, and the two particle solutions for
non-purely imaginary rapidities. The Bethe-Yang lines fits the TCSA points
within an error less then one percent.

\begin{figure}
\begin{centering}
\includegraphics[bb = 0 0 800 540, scale=0.5, type=eps]{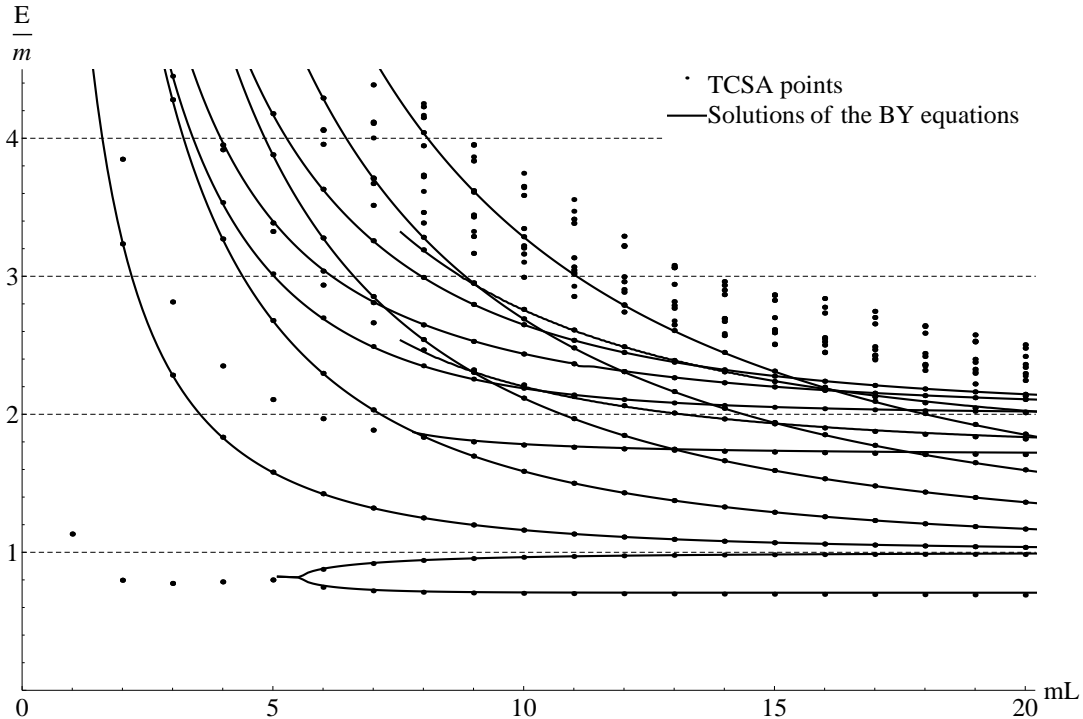}
\par\end{centering}

\caption{The real part of the  TCSA spectra together with the solutions of the Bethe-Yang equations
at $b=0.5$}

\end{figure}

If we choose $b=1.8$ we expect two defect bound state at infinite
volume. In finite volume, the corresponding states are the solutions
of the one-particle Bethe-Yang equations for purely imaginary rapidities.
These equations have solutions only for $mL>3$, for smaller volumes
the rapidities have non-zero real part. The corresponding TCSA points
are real for $mL>3$, but for smaller volumes these points become
complex. But for $mL<5$ the Bethe-Yang lines don't fit the TCSA
points because of the finite size corrections. We also solved the one-particle
Bethe-Yang equation for non-imaginary rapidities as well.

We can also identify the two-particle states solving the Bethe-Yang
equations. Similarly to the case of $b=0.5$ we have solutions where
none of the rapidities are imaginary, and solutions where one of them
is purely imaginary. Generally we have two solutions in the latter
case corresponding, in infinite volume, to one particle scattering
on an excited defect, but at $b=1.8$ we have two of them. These
equations for imaginary rapidity don't have a solution for every volume,
this also shows that in small volumes the Bethe-Yang equations are
not exact, and one should take into account the vacuum polarisation
effects. We plotted the energies of the solutions of the Bethe-Yang
equation only in that domain, where these solutions exist.

The energy lines of the solutions of the Bethe-Yang equations fits
the TCSA point within $4-5\,\%$ for volumes $mL>5$.

\begin{figure}
\begin{centering}
\includegraphics[bb = 0 0 800 540, scale=0.5, type=eps]{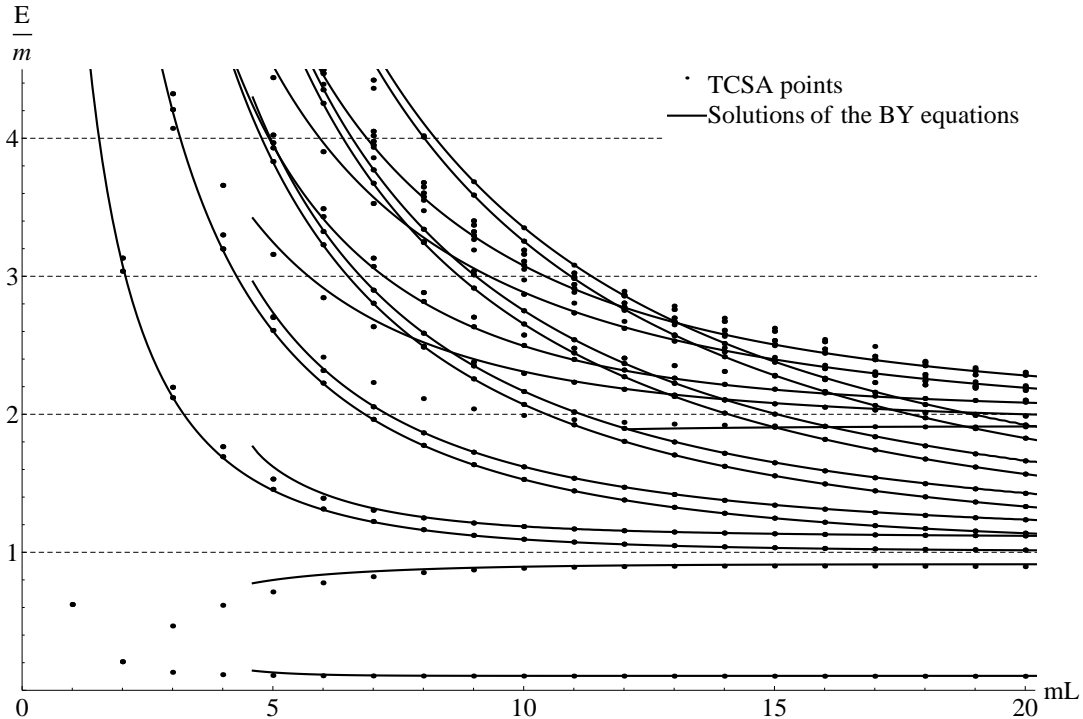}
\par\end{centering}

\caption{The real part of the TCSA spectra together with the solutions of the Bethe-Yang equations
at $b=1.8$. The ground state and the boundary bound-states are real. }

\end{figure}

\section{Conclusion}

We have carried out a detailed investigation of the integrable defects in the
scaling Lee-Yang model. Our approach is based on the perturbed CFT 
point of view. Thus, as a starting point, we solved the defect Lee-Yang model  
by calculating all of its structure constants. This is the first
defect conformal field theory solved at such an explicit level.  

We then determined the one parameter family of integrable perturbations  by
using defect conformal perturbation theory. Our findings, (\ref{eq:lmumu}), 
agree with the results of Runkel in \cite{Runkel:2010ym} obtained from an
alternative analysis.  

We matched the parameters of this UV description to the parameters of the 
IR scattering theory found in \cite{Bajnok:2007jg}
by fusing the defect to the boundary and using the boundary UV-IR relation
\cite{Dorey:1997yg}. 

We developed the defect truncated conformal space approach to calculate the 
finite size spectrum of the model and performed various 
numerical tests. In particular, we verified 
the UV-IR relation, the transmission factors and the bound-state spectrum
of  \cite{Bajnok:2007jg}. This was done by comparing the numerical spectrum
to the finite size correction determined by the Bethe-Yang equations. 
We also checked the defect energy contributions
and the leading Lüscher corrections to the vacuum energy.  
These provide convincing evidence for both our solution of the conformal 
defect Lee-Yang model and for the bootstrap results in  \cite{Bajnok:2007jg}. 

The Lee-Yang theory is a non-unitary theory, nevertheless its spectrum 
with periodic boundary condition is real. This is due to the PT-symmetry
of the model. Introducing defect perturbations we maintain this symmetry
but we obtained real spectrum only for real coupling constants. For 
purely imaginary defect perturbations only the ground state and the 
defect bound-states were real. This might be related to the fact 
that these states themselves are P-symmetric, contrary to the rest of the 
spectrum.

It is worth pointing out that although we write the chiral defect
fields as $\varphi$ and $\bar\varphi$ and their couplings as $\mu$ and
$\bar\mu$, the fields are actually real, self-conjugate fields and it
is no surprise that we only recovered a real spectrum for $\mu$ and
$\bar\mu$ both real. 

Our developments provide a firm basis to proceed with further work on 
the defect Lee-Yang model. For example, based
on the infinite volume defect form factors \cite{Bajnok:2009hp},  one could 
establish the theory of finite volume defect form factors. These 
results could then be checked directly by our DTCSA method. 
This will be the subject of a forthcoming paper.

It will also be interesting to investigate the full space of
non-integrable perturbations of the defect, both in the massless and
massive cases, using the DTCSA method. Previous investigations of
defect perturbations have been limited to the massless case
(see eg \cite{Kormos:2009sk}) and have yielded interesting results for the
space of RG flows including flows from purely transmitting defects to
purely reflecting defects. A similarly interesting picture is expected
for the space of RG flows in the massive Lee-Yang model.

The Lee-Yang model is the simplest conformal field theory, which 
we solved explicitly in the presence of a topological defect. Our 
analysis is quite general, however and can be easily generalized to any 
minimal model, as their topological defects are already classified
\cite{Petkova:2000ip}. These models then could be perturbed and the 
integrable perturbations classified.

\subsection*{Acknowledgements}

We thank Ferenc Wágner for collaboration at an early stage of the project, 
Gábor Takács and Francesco Buccheri for discussions.
GMTW would like to thank Ingo Runkel for discussions, ELTE for
hospitality while some of this research 
was carried out and STFC for partial support under grant 
ST/J002798/1. Z. Bajnok and L. Holló was supported by OTKA K81461
and by an MTA Lendület grant.

\appendix

\section{Structure constants of the defect Lee-Yang model}

In this section we solve the sewing relations for the defect conformal
Lee-Yang model and determine all the structure constants. Motivated by 
TCSA considerations we place the defect at $y=0$ and $x>0$ with 
$z=x+i y$. We start with the description of the relevant conformal blocks.  

The Virasoro algebra with $c= - \frac {22}{5}$  contains only one irredicble highest weight module with non-vanishing highest
weight $h=-\frac{1}{5}$. This module contains a singular vector at
level 2 
\begin{equation}
\left( L_{-1}^{2}-\frac{2}{5}L_{-2} \right) \vert h\rangle=0
\end{equation}
which leads to differential equations for the chiral correlations
functions (conformal blocks). Let us denote the chiral field with
weight $h$ by $\phi$. The matrix elements of $\phi(z)$ between
highest weight states have the following coordinate dependence:
\begin{equation}
\langle0\vert\phi(z)\vert0\rangle=0\quad;\quad\langle0\vert\phi(z)\vert h\rangle\propto z^{-2h}
\quad;\quad\langle h\vert\phi(z)\vert0\rangle\propto 1\quad;\quad\langle h\vert\phi(z)\vert h\rangle
\propto z^{-h}
\end{equation}
The matrix elements of $\phi(1)\phi(z)$ are proportional to 
\begin{equation}
  \langle0\vert\phi(1)\phi(z)\vert0\rangle\propto (1-z)^{-2h}
\; ;\quad
  \langle0\vert\phi(1)\phi(z)\vert h\rangle\propto(z(1-z))^{-h}
\; ;\quad
  \langle h\vert\phi(1)\phi(z)\vert0\rangle\propto (1-z)^{-h}
\end{equation}
finally from the decoupling of the singular vector we obtain a second
order hypergeometric differential equation, which can be solved
as 
\begin{equation}
\langle h\vert\phi(1)\phi(z)\vert h\rangle=c_{1}f_{1}(z)+c_{2}f_{2}(z)
\end{equation}
where 
\begin{equation}
  f_{1}(z)
= (z(1-z))^{-h}\,_{2}F_{1}({ \scriptstyle \frac 15,\frac{2}{5};\frac{4}{5}}\vert z)
\quad;\qquad 
  f_{2}(z)
= (z^{2}(1-z))^{-h}\,_{2}F_{1}({\scriptstyle \frac{2}{5},\frac{3}{5};\frac{6}{5}}\vert z)
\end{equation}
are the canonical solutions around $z\to0$, i.e. $f_{1}(z)=z^{-h}(1+a_{1}z+\dots)$
and $f_{2}(z)=z^{-2h}(1+a_{2}z+\dots)$. There is a canonical basis
around $z\to1$, too: 
\begin{equation}
\langle h\vert\phi(1)\phi(z)\vert h\rangle=\tilde{c}_{1}g_{1}(z)+\tilde{c}_{2}g_{2}(z)
\end{equation}
such that $g_{1}(z)=(1-z)^{-h}(1+\tilde{a}_{1}(1-z)+\dots)$ 
and $g_{2}(z)=(1-z)^{-2h}(1+\tilde{a}_{2}(1-z)+\dots)$.
As both are solutions of the same differential equations they can
be expressed in terms of each other as 
\begin{equation}
f_{i}(z)=\Gamma_{ij}g_{j}(z)
\end{equation}
with 
\begin{equation}
   \Gamma_{11}
= -\Gamma_{22}
=  \beta ^{-2}
\;,\;\;
   \Gamma_{12}
= -\beta^{-2} \alpha ^{-2} \;,\;\;
   \Gamma_{21}
= \alpha^2 
\end{equation}
where we use 
\begin{equation}
\alpha=\sqrt{\frac{\Gamma(\frac{1}{5})\Gamma(\frac{6}{5})}{\Gamma(\frac{3}{5})
\Gamma(\frac{4}{5})}}\quad;\quad\beta=\sqrt{\frac{2}{1+\sqrt{5}}}=\frac{1}{\sqrt{\eta+\eta^{-1}}}
\quad;\quad\eta=e^{\frac{i\pi}{5}}\quad;\quad\rho=e^{\frac{i \pi}{5}}
\end{equation}

\subsubsection*{Bulk structure constants}

The bulk operators are in a one-to-one correspondence with the bulk
Hilbert space: $V_{0}\otimes \bar{V}_{0}+V_{1}\otimes \bar{V}_{1}$ and the structure
constants can be calculated in this theory. Let us denote the $(h,h)$
field by $\Phi(z,\bar{z})$. It has the OPE
\begin{equation}
\Phi(z,\bar{z})\Phi(0,0)=C_{\Phi\Phi}^{\mathbb{I}}(z\bar{z})^{2/5}(\mathbb{I}+\dots)+
C_{\Phi\Phi}^{\Phi}(z\bar{z})^{1/5}(\Phi(0,0)+\dots)
\end{equation}
The four point function can be written in the two canonical bases
as 
\begin{equation}
\langle\Phi\vert\Phi(1,1)\Phi(z,\bar{z})\vert\Phi\rangle=a_{ij}f_{i}(z)f_{j}(\bar{z})=
\tilde{a}_{ij}g_{i}(z)g_{j}(\bar{z})
\end{equation}
 From the two different evaluation of the OPEs we can extract 
\begin{equation}
a_{11}=\tilde{a}_{11}=C_{\Phi\Phi}^{\mathbb{I}} \left( C_{\Phi\Phi}^{\Phi} \right)^{2}\quad;\quad a_{22}=
\tilde{a}_{22}=\left( C_{\Phi\Phi}^{\mathbb{I}} \right)^{2}
\end{equation}
Using the coefficient for the change of basis we obtain: 
\begin{equation}
C_{\Phi\Phi}^{\Phi}=\sqrt{-C_{\Phi\Phi}^{\mathbb{I}}}\alpha^{2}\beta
\end{equation}
As the three point function can be written as
\begin{equation}
\langle\Phi\vert\Phi(1,1)\vert\Phi\rangle=C_{\Phi\Phi\Phi}=C_{\Phi\Phi}^{\mathbb{I}}C_{\Phi\Phi}^{\Phi}
\end{equation}
 reality of $\Phi^{\dagger}=\Phi$ requires real $C_{\Phi\Phi\Phi}$.
We achieve this by choosing the normalization as 
\begin{equation}
C_{\Phi\Phi}^{\mathbb{I}}=-1\qquad;\qquad C_{\Phi\Phi}^{\Phi}=\alpha^{2}\beta=1.91131...
\end{equation}

\subsubsection*{Defect Hilbert space}

The defect Hilbert space is given by $V_{1}\otimes \bar{V}_{0}+V_{0}\otimes \bar{V}_{1}
+V_{1}\otimes \bar{V}_{1}$.
(It is like taking the fusion product of a chiral field with all bulk fields). 
The primary fields with weights $(h,0)$, $(0,h)$ and $(h,h)$ will
be denoted as $d$, $\bar{d}$ and $D$, respectively. We normalize
them as
\begin{equation}
\langle d\vert d\rangle=C_{dd}^{\mathbb{I}}=\langle\bar{d}\vert\bar{d}\rangle=
C_{\bar{d}\bar{d}}^{\mathbb{I}}=1 \quad ;\qquad \langle D\vert D\rangle=C_{DD}^{\mathbb{I}}=-1
\end{equation}
 and all other matrix elements are vanishing.

\subsubsection*{Defect operators}

The defect operators are in one-to-one correspondence to the Hilbert
space containing two defects: $V_{0}\otimes \bar{V}_{0}+V_{1}\otimes \bar{V}_{0}
+V_{0} \otimes \bar{V}_{1}+2\cdot V_{1}\otimes \bar{V}_{1}$.
(It is like taking the fusion product of a chiral field with the defect Hilbert space). 
The primary fields and weights are as follows: $\mathbb{I}$ with
$(0,0)$, $\varphi$ and $\bar{\varphi}$ with $(h,0)$ and $(0,h)$,
finally we have two fields $\Phi_{+}$ and $\Phi_{-}$ both with weights
$(h,h)$. We will choose them as the lower/upper limits of the bulk
field on the defect
\begin{equation}
\Phi_{\pm}(x)=\lim_{y\to\mp0}\Phi(z,\bar{z})\quad;\qquad z=x+iy
\end{equation}
As the map from the cylinder to the plane is $z=e^{-i \frac{2\pi}{L}\zeta}$
the left/right limit on the cylinder corresponded to the lower/upper limit on the plane. 
This implies the normalization of the fields
\begin{equation}
 C_{\Phi_{+}\Phi_{+}}^{\mathbb{I}}=C_{\Phi_{-}\Phi_{-}}^{\mathbb{I}}=C_{\Phi \Phi}^{\mathbb{I}} =-1
\end{equation} 
 In order to maintain reality of the chiral fields we normalize them as 
\begin{equation}
C_{\varphi\varphi}^{\mathbf{\mathbb{I}}}=C_{\bar{\varphi}\bar{\varphi}}^{\mathbf{\mathbb{I}}}=-1
\end{equation}
As the fields are real, complex conjugation
$z\leftrightarrow\bar{z}$ will make the changes: 
\begin{equation}
z\leftrightarrow\bar{z}\quad,\qquad\Phi_{\pm}\leftrightarrow\Phi_{\mp}\quad;
\qquad\varphi\leftrightarrow\bar{\varphi}
\end{equation}

\subsubsection*{Defect OPEs}

The defect operators have the following operator product expansions
\begin{eqnarray}
\varphi(z)\varphi(w) & = & C_{\varphi \varphi}^{\mathbb{I}}\vert z-w\vert^{2/5}+
C_{\varphi\varphi}^{\varphi}\vert z-w\vert^{1/5}\varphi(w)\dots\\
\varphi(z)\bar{\varphi}(w) & = & C_{\varphi\bar{\varphi}}^{\Phi_{+}}\Phi_{+}(w)+
C_{\varphi\bar{\varphi}}^{\Phi_{-}}\Phi_{-}(w)\dots\\ \nonumber
\\ 
\bar{\varphi}(z)\varphi(w) & = & C_{\bar{\varphi}\varphi}^{\Phi_{+}}\Phi_{+}(w)+
C_{\bar{\varphi}\varphi}^{\Phi_{-}}\Phi_{-}(w)\dots\\
\bar{\varphi}(z)\bar{\varphi}(w) & = & C_{\bar{\varphi}
\bar{ \varphi}}^{\mathbb{I}}
\vert z-w\vert^{2/5}+C_{\bar{\varphi}\bar{\varphi}}^{\bar{\varphi}}
\vert z-w\vert^{1/5}\bar{\varphi}(w)\dots\\ \nonumber
\\
\Phi_{+}(z)\Phi_{+}(w) & = & C_{\Phi_{+}\Phi_{+}}^{\mathbf{\mathbb{I}}}\vert z-w
\vert^{4/5}+C_{\Phi\Phi}^{\Phi} \vert z-w\vert^{2/5}\Phi_{+}(w)\dots\\ \nonumber 
\Phi_{+}(z)\Phi_{-}(w) & = & C_{\Phi_{+}\Phi_{-}}^{\mathbf{\mathbb{I}}}
\vert z-w\vert^{4/5}+C_{\Phi_{+}\Phi_{-}}^{\varphi}\vert z-w\vert^{3/5}\varphi(w)+
C_{\Phi_{+}\Phi_{-}}^{\bar{\varphi}}\vert z-w\vert^{3/5}\bar{\varphi}(w)+\\ 
 &  & +C_{\Phi_{+}\Phi_{-}}^{\Phi_{+}}\vert z-w\vert^{2/5}\Phi_{+}(w)+
C_{\Phi_{+}\Phi_{-}}^{\Phi_{-}}\vert z-w\vert^{2/5}\Phi_{-}(w)\dots\\ \nonumber
\\ \nonumber 
\Phi_{-}(z)\Phi_{+}(w) & = & C_{\Phi_{-}\Phi_{+}}^{\mathbf{\mathbb{I}}}\vert z-w\vert^{4/5}+
C_{\Phi_{-}\Phi_{+}}^{\varphi}\vert z-w\vert^{3/5}\varphi(w)+
C_{\Phi_{-}\Phi_{+}}^{\bar{\varphi}}\vert z-w\vert^{3/5}\bar{\varphi}(w)+\\ 
 & + & C_{\Phi_{-}\Phi_{+}}^{\Phi_{+}}\vert z-w\vert^{2/5}\Phi_{+}(w)+C_{\Phi_{-}
\Phi_{+}}^{\Phi_{-}}\vert z-w\vert^{2/5}\Phi_{-}(w)\dots\\ 
\Phi_{-}(z)\Phi_{-}(w) & = & C_{\Phi_{-}\Phi_{-}}^{\mathbf{\mathbb{I}}}\vert z-w\vert^{4/5}
+C_{\Phi\Phi}^{\Phi}\vert z-w\vert^{2/5}\Phi_{-}(w)\dots
\end{eqnarray}
where we have exploited the relation of $\Phi$ and $\Phi_{\pm}$
to write $C_{\Phi_{+}\Phi_{+}}^{\Phi_{+}}=C_{\Phi_{-}\Phi_{-}}^{\Phi_{-}}=C_{\Phi\Phi}^{\Phi}$.

\subsubsection*{Matrix elements of $\varphi$ and $\bar{\varphi}$}

The non-vanishing matrix elements of the defect operators at $z=1$
on the highest weight basis $(d,D,\bar{d})$ can be calculated 
from the matrix form of the OPEs
\begin{equation}
\hat{\varphi}=\left(\begin{array}{ccc}
C_{\varphi d}^{d} & 0 & 0\\
0 & C_{\varphi D}^{D} & C_{\varphi\bar{d}}^{D}\\
0 & C_{\varphi D}^{\bar{d}} & 0
\end{array}\right)\quad;\qquad\hat{\bar{\varphi}}=\left(\begin{array}{ccc}
0 & C_{\bar{\varphi}D}^{d} & 0\\
C_{\bar{\varphi}d}^{D} & C_{\bar{\varphi}D}^{D} & 0\\
0 & 0 & C_{\bar{\varphi}\bar{d}}^{\bar{d}}
\end{array}\right)
\end{equation}
In order to get the matrix elements we multiply with the normalization of states: 
$\langle i \vert \varphi \vert j \rangle =C_{\varphi j}^k C_{ik}^\mathbb{I}$. 
Since complex conjugation relates the two by changing $d\leftrightarrow\bar{d}$
we determine only the first. Analyzing carefully the OPEs we can express
the various matrix elements of $\varphi(1)\varphi(z)$ in terms of
the chiral blocks: 
\begin{equation}
\langle d\vert\varphi(1)\varphi(z)\vert d\rangle=\left(C_{\varphi d}^{d}\right)^{2}f_{1}(z)=
C_{\varphi\varphi}^{\mathbf{\mathbb{I}}}g_{2}(z)+C_{\varphi\varphi}^{\varphi}C_{\varphi d}^{d}g_{1}(z)
\end{equation}
\begin{equation}
-\langle D\vert\varphi(1)\varphi(z)\vert D\rangle=\left(C_{\varphi D}^{D}\right)^{2}f_{1}(z)
+C_{\varphi D}^{\bar{d}}C_{\varphi\bar{d}}^{D}f_{2}(z)=C_{\varphi\varphi}^{\mathbf{\mathbb{I}}}g_{2}(z)
+C_{\varphi\varphi}^{\varphi}C_{\varphi D}^{D}g_{1}(z)
\end{equation}
\begin{equation}
\langle\bar{d}\vert\varphi(1)\varphi(z)\vert\bar{d}\rangle=
C_{\varphi\bar{d}}^{D}C_{\varphi D}^{\bar{d}}(1-z)^{-2h}=
C_{\varphi\varphi}^{\mathbf{\mathbb{I}}}(1-z)^{-2h}
\end{equation}
\begin{equation}
\langle\bar{d}\vert\varphi(1)\varphi(z)\vert D\rangle=
C_{\varphi D}^{D}C_{\varphi D}^{\bar{d}}z^{-h}(1-z)^{-h}=
C_{\varphi\varphi}^{\varphi}C_{\varphi D}^{\bar{d}}z^{-h}(1-z)^{-h}
\end{equation}
\begin{equation}
-\langle D\vert\varphi(1)\varphi(z)\vert\bar{d}\rangle=
C_{\varphi\bar{d}}^{D}C_{\varphi D}^{D}(1-z)^{-h}=
C_{\varphi\varphi}^{\varphi}C_{\varphi\bar{d}}^{D}(1-z)^{-h}
\end{equation}
From which it easily follows that $C_{\varphi\bar{d}}^{D}
C_{\varphi D}^{\bar{d}}=C_{\varphi\varphi}^{\mathbf{\mathbb{I}}}=-1$
and $C_{\varphi D}^{D}=C_{\varphi\varphi}^{\varphi}$. Furthermore,
we found that 
\begin{equation}
C_{\varphi\varphi}^{\varphi}=C_{\varphi D}^{D}=\alpha\beta^{-1}\quad\quad C_{\varphi d}^{d}=\alpha\beta
\end{equation}
We can write the analogous equations by changing $d\leftrightarrow\bar{d}$
and $\varphi\leftrightarrow\bar{\varphi}$. The result is 
\begin{equation}
C_{\bar{\varphi}d}^{D}C_{\bar{\varphi}D}^{d}=C_{\bar{\varphi}\bar{\varphi}}^{\mathbf{\mathbb{I}}}=-1
\quad;\quad C_{\bar{\varphi}\bar{\varphi}}^{\bar{\varphi}}=C_{\bar{\varphi}D}^{D}=
\alpha\beta^{-1}\quad\quad C_{\bar{\varphi}\bar{d}}^{\bar{d}}=\alpha\beta
\end{equation}
 although we could have changed the sign of $\bar{\varphi}$ which
is still a solution.

\subsubsection*{Matrix elements of $\Phi_{\pm}$}

The matrix elements of $\Phi_{+}$ and $\Phi_{-}$ are related either
by complex conjugation or by analyzing the matrix elements of the
bulk field $\Phi(z,\bar{z})$ and taking the two limits $\theta\to0$
and $\theta\to2\pi$ in 
\begin{equation}
\langle i\vert\Phi(z,\bar{z})\vert j\rangle=\langle i\vert\Phi_{-}(1)\vert j
\rangle z^{h_{i}-h_{j}-h}\bar{z}^{\bar{h}_{i}-\bar{h}_{j}-h}\quad;\qquad z=re^{i\theta}
\end{equation}
This implies
\begin{equation}
\langle i|\Phi_{-}(1)|j\rangle=e^{2\pi i(h_{i}-\bar{h}_{i}-(h_{j}-\bar{h}_{j}))}\langle 
i\vert\Phi_{+}(1)\vert j\rangle\qquad;\qquad\xi=e^{i\frac{2\pi}{5}}
\end{equation}
The $\Phi_{\pm}(1)$ matrix elements can be parametrised as 
\begin{equation}
\hat{\Phi}_{-}=\left(\begin{array}{ccc}
0 & \ensuremath{C_{\Phi_{-}D}^{d}} & C_{\Phi_{-}\bar{d}}^{d}\\
\ensuremath{C_{\Phi_{-}d}^{D}} & C_{\Phi_{-}D}^{D} & C_{\Phi_{-}\bar{d}}^{D}\\
\ensuremath{C_{\Phi_{-}d}^{\bar{d}}} & C_{\Phi_{-}D}^{\bar{d}} & 0
\end{array}\right)\quad;\qquad\hat{\Phi}_{+}=\left(\begin{array}{ccc}
0 & \xi^{-1}\ensuremath{C_{\Phi_{-}D}^{d}} & \xi^{-2}C_{\Phi_{-}\bar{d}}^{d}\\
\xi\ensuremath{C_{\Phi_{-}d}^{D}} & C_{\Phi_{-}D}^{D} & \xi^{-1}C_{\Phi_{-}\bar{d}}^{D}\\
\xi^{2}\ensuremath{C_{\Phi_{-}d}^{\bar{d}}} & \xi C_{\Phi_{-}D}^{\bar{d}} & 0
\end{array}\right)
\end{equation}
These matrix elements can be determined from the correlation functions
\begin{equation}
\langle i\vert\varphi(1)\bar{\varphi}(z)\vert j\rangle=C_{i l}^{\mathbb{I}}
C_{\varphi k}^{l}
C_{\bar{\varphi}j}^{k}z^{\bar{h}_{i}-\bar{h}_{j}-h}=
C_{i l}^{\mathbb{I}}(
C_{\varphi\bar{\varphi}}^{\Phi_{-}}C_{\Phi_{-}j}^{l}+
C_{\varphi\bar{\varphi}}^{\Phi_{+}}C_{\Phi_{+}j}^{l})z^{\bar{h}_{i}-\bar{h}_{j}-h}
\end{equation}
In matrix notation they read as 
\begin{equation}
\hat{\varphi}\hat{\bar{\varphi}}=C_{\varphi\bar{\varphi}}^{\Phi_{-}}\hat{\Phi}_{-}+
C_{\varphi\bar{\varphi}}^{\Phi_{+}}\hat{\Phi}_{+}
\end{equation}
By solving the equations we found a one parameter family of solutions.
We fixed this freedom by choosing 
\begin{equation}
C_{\varphi D}^{\bar{d}}=C_{\bar{\varphi}D}^{d}=1\quad;\qquad C_{\varphi\bar{d}}^{D}=C_{\bar{\varphi}d}^{D}=-1
\end{equation}
The rest of the coefficients are 

\begin{equation}
C_{\Phi_{+}\Phi_{+}}^{\Phi_{+}}=C_{\Phi_{-}\Phi_{-}}^{\Phi_{-}}=\alpha^{2}\beta \quad;\qquad C_{\varphi d}^{d}=C_{\bar{\varphi}\bar{d}}^{\bar{d}}=\alpha\beta
\end{equation}
\begin{equation} C_{\varphi\bar{\varphi}}^{\Phi_{-}}=C_{\bar{\varphi}\varphi}^{\Phi_{+}}=\frac{\beta^{-1}}{1+\eta^{-1}}=\frac{\rho}{\sqrt[4]{5}}  \quad;\qquad C_{\varphi\bar{\varphi}}^{\Phi_{+}}=C_{\bar{\varphi}\varphi}^{\Phi_{-}}=\frac{\beta^{-1}}{1+\eta}=\frac{\rho^{-1}}{\sqrt[4]{5}}
\end{equation}
\begin{equation}
C_{\Phi_{-}D}^{d}=C_{\Phi_{+}D}^{\bar{d}}=\alpha\eta \quad;\qquad C_{\Phi_{-}D}^{\bar{d}}=C_{\Phi_{+}D}^{d}=\alpha\eta^{-1}
\end{equation}
\begin{equation}
C_{\Phi_{-}\bar{d}}^{d}=C_{\Phi_{+}d}^{\bar{d}}=-\eta^{2}\beta^{-1} \quad;\qquad C_{\Phi_{-}d}^{\bar{d}}=C_{\Phi_{+}\bar{d}}^{d}=-\eta^{-2}\beta^{-1}
\end{equation}
\begin{equation}
C_{\Phi_{-}d}^{D}=C_{\Phi_{+}\bar{d}}^{D}=-\eta^{-1}\alpha \quad;\qquad C_{\Phi_{-}\bar{d}}^{D}=C_{\Phi_{+}d}^{D}=-\eta\alpha
\end{equation}
\begin{equation}
C_{\Phi_{-}D}^{D}=C_{\Phi_{+}D}^{D}=\alpha^{2}\beta^{-1}
\quad;\qquad C_{\Phi_{-}\Phi_{+}}^{\mathbb{I}}=C_{\Phi_{+}\Phi_{-}}^{\mathbb{I}}=\left(1+\beta^{-2}\right)
\end{equation}
\begin{equation}
C_{\Phi_{-}\Phi_{+}}^{\Phi_{-}}=C_{\Phi_{+}\Phi_{-}}^{\Phi_{+}}=C_{\Phi_{-}\Phi_{+}}^{\Phi_{+}}=C_{\Phi_{+}\Phi_{-}}^{\Phi_{-}}=\beta^{-1}\alpha^{2}
\end{equation}
\begin{equation}
C_{\Phi_{-}\Phi_{+}}^{\varphi}=C_{\Phi_{+}\Phi_{-}}^{\bar{\varphi}}=-\rho \alpha \beta^{-2} \sqrt[4]{5} 
\quad;\quad C_{\Phi_{-}\Phi_{+}}^{\bar{\varphi}}=C_{\Phi_{+}\Phi_{-}}^{\varphi}=-\rho^{-1} \alpha \beta^{-2} \sqrt[4]{5} 
\end{equation}
\begin{equation}
C_{\Phi_{-}\bar{\varphi}}^{\varphi}=C_{\Phi_{+}\varphi}^{\bar{\varphi}}=C_{\bar{\varphi}\Phi_{+}}^{\varphi}=C_{\varphi\Phi_{-}}^{\bar{\varphi}}=-\beta^{-1}\eta 
\end{equation}
\begin{equation}
C_{\bar{\varphi}\Phi_{-}}^{\varphi}=C_{\varphi\Phi_{+}}^{\bar{\varphi}}=C_{\Phi_{+}\bar{\varphi}}^{\varphi}=C_{\Phi_{-}\varphi}^{\bar{\varphi}}=-\beta^{-1}\eta^{-1}
\end{equation}
\begin{equation}
C_{\Phi_{-}\varphi}^{\Phi_{-}}=C_{\Phi_{+}\bar{\varphi}}^{\Phi_{+}}=C_{\bar{\varphi}\Phi_{-}}^{\Phi_{-}}=C_{\varphi\Phi_{+}}^{\Phi_{+}}=\frac{\alpha}{2}\left(\left(\beta+\beta^{-1}\right)-i\frac{1}{\sqrt[4]{5}}\right)=\frac{\alpha}{2} \rho^{-1} \sqrt{\frac{6}{\sqrt{5}}+2}
\end{equation}
\begin{equation}
C_{\Phi_{-}\bar{\varphi}}^{\Phi_{-}}=C_{\Phi_{+}\varphi}^{\Phi_{+}}=C_{\varphi\Phi_{-}}^{\Phi_{-}}=C_{\bar{\varphi}\Phi_{+}}^{\Phi_{+}}=\frac{\alpha}{2}\left(\left(\beta+\beta^{-1}\right)+i\frac{1}{\sqrt[4]{5}}\right)=\frac{\alpha}{2} \rho \sqrt{\frac{6}{\sqrt{5}}+2}
\end{equation}
\begin{equation}
C_{\varphi\Phi_{+}}^{\Phi_{-}}=C_{\bar{\varphi}\Phi_{-}}^{\Phi_{+}}=C_{\Phi_{+}\bar{\varphi}}^{\Phi_{-}}=C_{\Phi_{-}\varphi}^{\Phi_{+}}=\frac{\alpha\beta}{2}\left(1+i\left(\beta-\beta^{-1}\right)\frac{1}{\sqrt[4]{5}}\right)=\frac{\alpha}{2} \rho^{-1} \sqrt{\frac{6}{\sqrt{5}}-2}
\end{equation}
\begin{equation}
C_{\bar{\varphi}\Phi_{+}}^{\Phi_{-}}=C_{\varphi\Phi_{-}}^{\Phi_{+}}=C_{\Phi_{+}\varphi}^{\Phi_{-}}=C_{\Phi_{-}\bar{\varphi}}^{\Phi_{+}}=\frac{\alpha\beta}{2}\left(1-i\left(\beta-\beta^{-1}\right)\frac{1}{\sqrt[4]{5}}\right)=\frac{\alpha}{2} \rho \sqrt{\frac{6}{\sqrt{5}}-2}
\end{equation}

In matrix notation
\begin{equation}
\hat{\Phi}_{-}=\left(\begin{array}{ccc}
0 & \ensuremath{\eta \alpha} & -\eta^{2}\beta^{-1}\\
\ensuremath{-\eta^{-1}\alpha} & \alpha^{2}\beta^{-1} & \ensuremath{-\eta^{1}\alpha}\\
\ensuremath{-\eta^{-2}\beta^{-1}} & \eta^{-1}\alpha & 0
\end{array}\right)\; ;\quad\hat{\Phi}_{+}=\left(\begin{array}{ccc}
0 & \ensuremath{\eta^{-1} \alpha} & -\eta^{-2}\beta^{-1}\\
\ensuremath{-\eta^{1}\alpha} & \alpha^{2}\beta^{-1} & \ensuremath{-\eta^{-1}\alpha}\\
\ensuremath{-\eta^{2}\beta^{-1}} & \eta^{1}\alpha & 0
\end{array}\right)
\end{equation}

\section{Perturbation theory calculations.}
\label{app:details}

\newcommand{\e}{\epsilon}
\newcommand{\vac}{{|0\rangle}}
\newcommand{\cev}[1]{{\langle{#1}|}}
\newcommand{\vecc}[1]{{|{#1}\rangle}}
\newcommand{\rd}{{\mathrm d}}

In this Appendix we calculate the bulk-defect operator expansion induced by
simultaneous bulk,  chiral and anti-chiral defect perturbations.
From this we can easily calculate the jump in $T$ and $\bar T$ across
the defect.

 Operator equations are local, which are understood within
correlators in the perturbed theory. This means we require them in the weak
sense for any of their matrix elements. For technical reasons we present the 
calculation here for matrix elements in the theory where two defect lines
are included, i.e. when there is a one-to-one correspondence between 
defect operators and vectors of the Hilbert space. We place the defect
at $ x=0 $ and sometimes write out explicitly that fields depend on 
$z=x+iy$, such as like $\varphi(iy)$. 

There are singularities in the perturbative expansion of correlation
functions including $T(0)$ coming from integration over
the boundary perturbation. The solution is to consider instead the
regularised field 
\begin{equation}
T_R(0) = T(0) + \frac{a \mu}\e \varphi(0)\;.
\end{equation}
The constant $a$ can be
fixed by requiring the bulk-defect OPE of $T(x)$ to be
\begin{equation}
 e^{-\delta S} T(x) = e^{-\delta S} T_R(0) + O(x,\mu)
\;.
\end{equation}
If we sandwich this identity between $\cev\varphi$ and $\vac$, we get
\begin{equation}
  \cev\varphi \,e^{-\delta S}\, (T(x) - T(0)) \vac
= ({a\mu}/\e) \cev\varphi \,e^{-\delta S}\,\varphi(0) \vac 
\;.
\end{equation}
We fix $a$ by differentiating both sides with respect to $\mu$ and
setting $\mu$ to zero. 
Since the singularity arises for $\e\to 0$, we can always take
$x>\epsilon$, to get 
\begin{eqnarray*}
  a
&=&  \e\lim_{R\to\infty}\int_{-R}^R 
\frac{\cev\varphi \varphi(iy)(T(x)-T(0)) \vac  }{\cev\varphi \varphi(0)\vac}
    dy\Big|_{\hbox{\tiny cutoff}}
\\
&=&  \e\lim_{R\to\infty}\int_{-R}^R
  \left( \frac{h}{(x-iy)^2}\theta(x^2+y^2-\e^2) -
  \frac{h}{(iy)^2}\theta(y^2-\e^2) \right) dy 
\\
&=& - h\e \lim_{R\to\infty} \left(
  \int_{-R}^R \frac{dy}{(y + ix)^2} 
- \int_{-R}^{-\epsilon}\frac{dy}{y^2}
- \int_\epsilon^R\frac{dy}{y^2}  
  \right)
\\&=& 
 {2h}
\;,
\end{eqnarray*}
where ``cutoff'' means we need to implement the short-distance cutoff
in the perturbative integrals. 

Having defined the renormalized field we can now write the general bulk-defect operator expansion for the
combined chiral, anti-chiral and bulk perturbations in the case ${x}>0$:
\begin{equation}
 T({x})
= T_R(0)
+ b^+\mu\partial_y\varphi(0)
+ \eta^+\lambda\Phi_+(0)
+ \eta^-\lambda\Phi_-(0)
+ \mu\bar\mu\Big(\rho\bar\varphi(0)\varphi(0)
+ \sigma\varphi(0)\bar\varphi(0)\Big)
+ \ldots
\end{equation}
We will need to find $b^+$, the coefficient of $\partial_y\varphi(iy) = i
\partial\varphi(iy)$.  
We have $\partial_y\varphi(0)\vac = iL_{-1}\vecc{\varphi}$ and so we find
$b^+$ by sandwiching 
\[
  T({x}) = T_R(0) + \cdots + \gamma\mu\partial_y\varphi(iy) + \cdots
\;,
\] 
between $\cev\psi = -\frac{i}{2h}\cev\varphi L_1$ and $\vac$,
differentiating with respect to $\mu$ and setting $\mu\to 0$, giving
\begin{eqnarray*}
b^+ 
&=& -\frac{i}{2h}\int
{  \cev\varphi L_1 \varphi(iy) (T({x}) - T(0)) \vac } / 
{\cev\varphi \varphi \rangle}\;
\rd y
\Big|_{\hbox{\tiny cutoff}}
\\
&=& -i \lim_{R\to\infty} \int_{y=-R}^R
  \left[ 
 \left(\frac{h{x}}{({x}-iy)^2} + \frac{(1-h)}{{x}-iy}\right)\theta(
 |{x}-iy|-\e)
- \frac{(1-h)}{-iy}\theta( |y|-\e)
  \rd y \right]
\\
&=&  -\lim_{R\to\infty} 
  \left[\int_{-R}^{-\e}+\int_{\e}^R (1-h)\frac{\rd y}{y}\right]
- \left[\int_{-R}^R \left( \frac{ih{x}}{(y+i{x})^2} +
  \frac{1-h}{y+i{x}}\right)\rd y\right]
\\
&=&  -\lim_{R\to\infty} \left(
 -(1-h)\log\left( \frac{i{x} + R}{i{x} - R} \right)
+ ih{x}\left[\frac{1}{R+i{x}}+\frac{1}{R-i{x}}\right]
\right)
\\
&=& -i\pi(1-h)
\;.
\end{eqnarray*}
We will also need $\eta^+$, the coefficient of $\Phi_+$ appearing in the
bulk-defect OPE. This can  be found by sandwiching the bulk-defect
operator expansion between
$\cev{+}$ and $\vac$, 
where the state $\cev{+}$ picks out the contribution from $\Phi(x)$
with positive $x$ ---
\[
 \cev{+} \Phi(x) \vac = 1\,\, (x>0)
\;,\;\;\;
 \cev{+} \Phi(x) \vac = 0\,\, (x<0)
\;.
\]
This state exists so long as $\Phi(x)$ is discontinuous across the
defect.
The result is
\[
 \eta^+ 
= \lim_{R\to\infty} \int_U \rd^2u \cev+ \Phi(u) (T({x}) - T(0)) \vac
\Big|_{\hbox{\tiny cutoff}}
= \lim_{R\to\infty} \int_{U_1} \rd^2u \frac{h}{(u-{x})^2}
 - \int_{U_2} \rd^2u \frac{h}{u^2}
\;,
\]
where the integration region $U = \{ u \in \mathbb{C} | Re(u)>0, |u|<R\}$,
$U_1 =  \{ u \in \mathbb{C} | Re(u)>0, |u|<R, |u-x|>\e\}$ and
$U_2 =  \{ u \in \mathbb{C} | Re(u)>0, \e<|u|<R \}$.
The second integral is zero, as can be found by taking $u=re^{i\theta}$. 
For the first integral, we can take $u={x} + re^{i\theta}$, and then
the integration region is approximately given by 
$U_3 = \{r e^{i\theta} | \e<r<R, r\cos\theta+x>0\}$.
The difference between this approximate region and the correct region
goes to zero as $R$ goes to infinity. 
We then find, with $\theta_0 = \sin^{-1}(x/R)$,
\begin{eqnarray*}
&&  \int_{U_2} \rd^2 u \frac{1}{(u-x)^2}
\simeq
   \int_{U_3} \frac{\rd r\,\rd\theta}{r e^{2i\theta}}
\\
&=&  \int_{r=\e}^R \int_{\theta=-\pi/2-\theta_0}^{\pi/2+\theta_0}
    \frac{\rd r\,\rd\theta}{r e^{2i\theta}}
 + \int_{\theta=-\pi}^{-\pi/2-\theta_0}
    \int_{r=\e}^{-x/\cos\theta}
     \frac{\rd r\,\rd\theta}{r e^{2i\theta}}
 + \int_{\theta=\pi/2+\theta_0}^{\pi}
    \int_{r=\e}^{-x/\cos\theta}
    \frac{\rd r\,\rd\theta}{r e^{2i\theta}}
\\
&=& 
   \int_{\theta=-\pi/2-\theta_0}^{\pi/2+\theta_0} e^{-2i\theta}\log(R/\e)\rd\theta
 + \int_{\theta=-\pi}^{-\pi/2-\theta_0}
    e^{-2i\theta}\log\left(-\frac x{\e\cos\theta}\right)\rd\theta 
\\
\nonumber&& + \int_{\theta=\pi/2+\theta_0}^{\pi}
    e^{-2i\theta}\log\left(-\frac x{\e\cos\theta}\right)\rd\theta 
\\
&=&
 \frac 12\left(
 2\theta_0 + \sin(2\theta_0) - \pi\right)
\end{eqnarray*}
In the limit $R\to\infty$, $\theta_0 \to 0$ and so 
$
 \eta^+ = - {h \pi}/2
\;.
$
We can likewise find the remaining coefficients to get the bulk-defect
operator expansion, valid for $x>0$,
\begin{equation}
 T({x})
= T_R(0)
-i\pi(1-h) \mu\partial_y\varphi(0)
-\frac{h\pi}2 \lambda(\Phi_+(0) - \Phi_-(0))
+ i\pi h \mu\bar\mu\Big[\varphi(0),\bar\varphi(0)\Big]
+ \ldots
\end{equation}
and so we find the jump in $T$ to be
\begin{eqnarray}
 \Delta T({x})
&=& 
2\pi i(1{-}h) \mu\,\partial_y\varphi(0)
- {h\pi}\lambda\,\Delta\Phi(0)
-  2\pi i h \mu\bar\mu\Big[\varphi(0),\bar\varphi(0)\Big]
+ \ldots
\\
\noalign{\noindent and likewise}
\nonumber \\[-15pt]
 \Delta \bar T({x})
&=& 
-2\pi i(1{-}h)\bar\mu\,\partial_y\bar\varphi(0)
- {h\pi}\lambda\,\Delta\Phi(0)
-  2\pi i h \mu\bar\mu\Big[\varphi(0),\bar\varphi(0)\Big]
+ \ldots
\end{eqnarray}

\newpage
\bibliographystyle{unsrt}

\end{document}